\documentclass[aps,pra,twocolumn,groupedaddress,superscriptaddress,bibnotes,amsfonts,longbibliography,citeautoscript,a4paper,10pt]{revtex4-2}
\usepackage{latexsym}
\usepackage{amsmath}
\usepackage{amssymb}
\usepackage{graphicx}
\usepackage[T1]{fontenc}
\usepackage[open]{bookmark}
\usepackage[english]{babel}

\usepackage{hyperref}
\usepackage{orcidlink}
\hypersetup{colorlinks=true,allcolors=blue}
\usepackage{xcolor} 
\begin{document}

\title{Structural Colours with Transition Metal Dichalcogenide Nanostructures}

\author{Ida Juliane Bundgaard\,\orcidlink{0009-0008-3424-0949}}
\email{bundgaard@mci.sdu}
\affiliation{POLIMA---Center for Polariton-driven Light--Matter Interactions, University of Southern Denmark, Campusvej 55, DK-5230 Odense M, Denmark}

\author{Catarina G. Ferreira\,\orcidlink{0000-0002-4088-1952}}
\affiliation{POLIMA---Center for Polariton-driven Light--Matter Interactions, University of Southern Denmark, Campusvej 55, DK-5230 Odense M, Denmark}
\affiliation{SCC --- SDU Climate Cluster, University of Southern Denmark, Campusvej 55, DK-5230 Odense M, Denmark}

\author{Yonas Lebsir\,\orcidlink{0000-0002-9383-0278}}
\affiliation{POLIMA---Center for Polariton-driven Light--Matter Interactions, University of Southern Denmark, Campusvej 55, DK-5230 Odense M, Denmark}

\author{Christos Tserkezis\,\orcidlink{0000-0002-2075-9036}}
\email{ct@mci.sdu.dk}
\affiliation{POLIMA---Center for Polariton-driven Light--Matter Interactions, University of Southern Denmark, Campusvej 55, DK-5230 Odense M, Denmark}

\date{\today}

\begin{abstract}
We introduce transiton metal-dichalcogenide (TMD) nanostructures as a promising platform for the realisation of structural colours. Processing of semianalytically calculated reflectance spectra of TMD nanosphere arrays shows a wide range of colours, which are obtained simply through tailoring the radius and separation of spheres in the array, with the size-dependent Mie modes of the nanoparticles being the primary contributor to the spectra. Additionally, it is demonstrated that further coverage of the colour space can be obtained by employing different materials or different lattice unit cells. Theoretical examination of the impact of the excitonic attributes of TMDs on the resulting structural colours indicates that self-hybridisation between nanoparticle modes and excitonic transitions may be employed for further tuneability. Moreover, the impact of TMD anisotropy on the structural colours is shown to be negligible for small structures at typical viewing angles, while the viewing angle itself may impact the colour. This work sets out to be a general investigation of TMD nanoarchitectures, with a focus on nanosphere arrays, for structural colours, by examining both inherent material features through the lens of colourimetry, and the ability of such structures to sustain a broad range of hues.
\end{abstract}

\maketitle

\section{Introduction}
\label{sec:introduction}
The colours in the world around us have long been fascinating to scientists; these may arise from inherent chemical properties of a material, causing bands of absorption and reflection, like in paints based on pigments, or from the structure of the object creating resonances to the the same end~\cite{Daqiqeh2021NanoStructuralColours}, like the colouration of butterflies\cite{Kinoshite2002MorphoButterfly}. The latter has been of increasing interest in the last century, due to the growing potential for exact manipulation of transmission and reflection spectra and driven in part by the continued advances in material science and fabrication technology. This has also motivated application driven research; here the focus may be on sustainability, as structural colours may present a route for more easily recyclable coloured products~\cite{Clausen2014ColourationPlastic} or photovoltaics which can be integrated directly into colourful building facades~\cite{Ferreira2025AngularColourIntegratedPV, Ferreira2026ColourPVOptimization,Zhang2019ColourSolarCellsLID,Kim2023ColoursPVDiffractive}. However, the focus can also be on security, where colour patterning can be used for information encryption~\cite{Lapidas2024PlasmonicEncryption, Syubaev2022SecurityLabelInfoEncryption, Zhou2024PolarisationSensitiveEncryption} and effective identification or anti-counterfeiting tools~\cite{Wang2021Nanopillars, Liu2019ChiralColour, HongNg2019Microtags, Hu2018LaserSteganography}. The choice of structure for each of these applications will differ according to the varying requirements; this may include considerations for spatial resolution, colour vibrancy, viewing independence (denoting whether the colour is consistent independent of the viewing angle), durability, and manufacturability\cite{Kristensen2016PlasmonColourGen}.

Structural colours can be obtained in a number of ways -- much attention has been focused on plasmonic systems, where the colours arise due to highly confined plasmonic resonances which may be combined with other structural contributions~\cite{Kristensen2016PlasmonColourGen, Cheng2015PlasmonColourPerfectAbsorption, Goh20143DStereoscopicPlasmonPrint, Franklin2020SelfAssemblyPlasmonicColours}. These systems are popular due to their strong collective resonances which lead to high field confinement~\cite{Gramotnev2010Plasmonics,Gramotnev2014Nanofocusing} and due to the ability for resolution beyond the diffraction limit~\cite{Kumar2012PrintingDiffractionLimit,Roberts2014SubwavelengthPrint}
that characterises them. Dielectrics, while lacking the strong and highly confined plasmon resonances, still have potential for structural colours~\cite{Kristensen2016PlasmonColourGen,Daqiqeh2021NanoStructuralColours}. The primary mechanism is Mie resonances which arise as somewhat localised resonances in dielectric nanoparticles (NPs)~\cite{todisco2020EPolaritonsSiDisk, Verre2019TMDMieNanodisc}, often called Mie resonators~\cite{evlyukhin_nl12,kivshar_nl22}. In high-index dielectrics (HIDs), with low intrinsic losses, these resonances may be quite strong~\cite{Arseniy2016DielectricResonance}, and thus can contribute to vibrant colours\cite{Cao2010SiliconWireColour, Faluraud2017SiliconFullColourPrints,Yang2020StructuralColoursSi3N4, Yang2020DielectricHighPerformance}.

Transition metal dichalcogenides (TMDs) present themselves as a promising candidate for the generation of structural colours due to their high refractive indices; furthermore, their current popularity in optoelectronic research~\cite{Manzeli20172DTMD, Zhao2026ExcitonPolaritonDiodes} opens the door to easy integration of colours in such devices. Recent interest stems largely from their excitonic properties and the possibility to use them as two-dimensional (2D) materials; due to their atomic structure and the weak van der Waals interactions between consecutive layers, it is possible to create sheets consisting of just one layer of the material~\cite{Manzeli20172DTMD}
and thus push miniaturisation of optoelectronic devices to its limits.
Another interesting property of TMDs is that they are generally uniaxially anisotropic~\cite{Ermolaev2021GiantAnisotropy,Neupane2019Anisotropy, Munkhbat2022Anisotropy}, meaning that the permittivity components along two axes are the same but differ from the third; the two identical permittivities may be called the ordinary direction or in-plane component (denoted $\varepsilon_\mathrm{o}$) and the different permittivity may be called the extraordinary direction or out-of-plane component (denoted $\varepsilon_\mathrm{e}$).
This combination of properties makes TMDs excellent candidates for many different applications; some which utilise the absorption of their excitonic transitions~\cite{Wang2016ExcitonCouplingWS2,Munkhbat2019PlexcitonsTMD} and others which need the low-loss properties in other regions of the electromagnetic spectrum~\cite{Munkhbat2023NTMD,Datta2020LowlossPlatform}.

In this work, TMD nanostructures are presented as a potential framework for the realisation of structural colours based upon a broader examination of arrays of spherical NPs and general TMD properties. The main focus is on structures made from tungsten disulfide (WS$_2$), primarily in the form of square arrays of identical NPs, utilising their Mie resonances as the primary optical response. While spherical geometries may not be the most feasible in terms of fabrication, they offer a first approximation of more complex morphologies, with spectra that can be calculated using semianalytical means, thus serving as a proof of concept for colour generation utilising TMD NPs. Initially, an investigation into the excitonic properties of the TMDs is undertaken, to determine the significance of excitonic transitions on the reflectance spectrum and, more importantly, the resulting colours. Due to the possible qualms of anisotropy, this aspect is also investigated; to keep this consideration purely analytical, thin anisotropic slabs are used, but any conclusions drawn on this type of structure should apply to other configurations as well. The pièce de résistance is the analysis of reflectance spectra from square arrays of spherical NPs with varying radii spanning from 30~nm to 150~nm and lattice constants spanning from the diameter of the particles to 800~nm and the presentation of the resulting colours, their viewing independence and their coverage of the conventional colour space. This also leads to discussions of further tailorability through the use of different variable parameters or different materials. This thorough investigation constitutes another volume in our library of potential materials and architectures for the realisation of structural colours.

\section{Methods}
\label{sec:methods}
In this paper, two primary nano-architectures are considered: slabs and arrays of spherical NPs. The methodology for calculating the spectra for each of the structures is different and described below in separate subsections. The method for calculating the corresponding colours, as described in a subsequent section, is the same regardless of the architecture. In this work, we focus on the colour resulting from reflection, corresponding to how the colours of most objects are observed (with the exception of stained glass windows and back-lit displays).
\subsection{Reflectance of uniaxially anisotropic slabs}
Perhaps, the simplest way to achieve structural colours is to exploit optical diffraction from slabs of a given material -- here slabs denote architectures where in the $xy$-plane the material stretches infinitely, but in the $z$-direction it is finite and the thickness of the material is still such that its optiucal properties are well described by bulk parameters.
The general reflection coefficient for a finite slab is given by \cite{BornWolfPrinciplesOptics}
\begin{equation}
    R=\frac{r_{12}+r_{23}\mathrm{e}^{2\mathrm{i}T k_{2z}}}{1+r_{12}r_{23}\mathrm{e}^{2\mathrm{i}Tk_{2z}}},
    \label{eq: Reflectance_Slab}
\end{equation}
where $r_{ij}$ represents the reflection coefficient at the interface between material $i$ and $j$, $T$ is the thickness of the slab and $k_{iz}$ is the wavevector normal to the interface in the corresponding medium.\\
Ordinarily, we use the Fresnel coefficients of reflection at an interface $r_{ij}$ given by~\cite{NovotnyPrinciplesNanoOptics}, 
\begin{equation}
    r^{(TE)}_{ij}=\frac{\mu_jk_{iz}-\mu_ik_{jz}}{\mu_jk_{iz}+\mu_ik_{jz}},\qquad
    r^{(TM)}_{ij}=\frac{\varepsilon_jk_{iz}-\varepsilon_ik_{jz}}{\varepsilon_jk_{iz}+\varepsilon_ik_{jz}},
    \label{eq: Fresnel coefficients}
\end{equation}
to calculate the reflection of the slab, where $\mu_{i}$  and $\varepsilon_{i}$ is the relative permeability and permittivity of medium $i$ respectively. Note here that we have introduced the superscripts TE and TM which represent the transverse electric and transverse magnetic polarisation of light, respectively. These equations do not, however, consider that the medium on either side of the interface may be anisotropic. TMDs are generally uniaxially anisotropic and as such their permittivity and permeability should be written as tensors \cite{NovotnyPrinciplesNanoOptics},\\
\begin{equation*}
    \varepsilon =
    \begin{pmatrix}
        \varepsilon_\mathrm{o}& & \\
         &\varepsilon_\mathrm{o}& \\
         & &\varepsilon_\mathrm{e}    
    \end{pmatrix},\qquad\qquad     
    \mu =
    \begin{pmatrix}
        \mu_\mathrm{o}& & \\
         &\mu_\mathrm{o}& \\
         & &\mu_\mathrm{e}    
    \end{pmatrix};
     \label{eq: Tensorial}
\end{equation*}
here, the optical axis has been chosen to be along $z$ and what follows the $k$-vector is assumed to have only $z$ and $x$ components. Note that for most optical materials, including the ones investigated in this work, the relative permeability is unity.
It can be shown that, when one medium is uniaxial, the reflection coefficients at each interface should be written as
\begin{equation}
    r^{(\mathrm{TE})}_{12}=\frac{\mu_{2\mathrm{o}} k_{1z}-\mu_1 k_{2z}}{k_{2z}\mu_1+k_{1z}\mu_{2\mathrm{o}}},\qquad r^{(\mathrm{TM})}_{12}=\frac{\varepsilon_{2\mathrm{o}}k_{1z}-\varepsilon_1 k_{2z}}{\varepsilon_{2\mathrm{o}} k_{1z} + \varepsilon_1 k_{2z}}
    \label{eq: Anisotropic coefficients 1}
\end{equation}
and
\begin{equation}
    r^{(\mathrm{TE})}_{23}=\frac{\mu_{3} k_{2z}-\mu_{2\mathrm{o}} k_{3z}}{\mu_{3} k_{2z}+\mu_{2\mathrm{o}} k_{3z}},
    \qquad
    r^{(\mathrm{TM})}_{23}=\frac{\varepsilon_{3}k_{2z}-\varepsilon_{2\mathrm{o}} k_{3z}}{\varepsilon_3 k_{2z} + \varepsilon_{2\mathrm{o}} k_{3z}}
    \label{eq: Anisotropic coefficients 2}
\end{equation}
where the second medium has been assumed uniaxial while media 1 and 3 are considered isotropic. The expressions in eq.~\eqref{eq: Anisotropic coefficients 1} and eq.~\eqref{eq: Anisotropic coefficients 2} can readily be substituted into eq.~\eqref{eq: Reflectance_Slab} to calculate the reflectance of an anisotropic slab, similarly to how eq.~\eqref{eq: Fresnel coefficients} is used for the isotropic slab.

The wave vector in the uniaxial medium ($k_{2z}$) can be determined using the similarly derived dispersion relations for both TE
\begin{equation}
    \frac{\omega^2}{c^2}=\frac{k_{x}^2}{\varepsilon_{\mathrm{o}}\mu_{\mathrm{e}}}+\frac{k_{z\mathrm{(TE)}}^2}{\varepsilon_{\mathrm{o}}\mu_{\mathrm{o}}}\quad\Rightarrow\quad k_{z\mathrm{(TE)}}=\sqrt{k_0^2\varepsilon_{\mathrm{o}}\mu_{\mathrm{o}}-k_x^2\mu_{\mathrm{o}}}
    \label{eq: Dispersion TE}
\end{equation}
and TM light
\begin{equation}
    \frac{\omega^2}{c^2}=\frac{k_{x}^2}{\varepsilon_{\mathrm{e}}\mu_{\mathrm{o}}}+\frac{k_{z\mathrm{(TM)}}^2}{\varepsilon_{\mathrm{o}}\mu_{\mathrm{o}}}\quad\Rightarrow\quad k_{z\mathrm{(TM)}}=\sqrt{k_0^2\varepsilon_{\mathrm{o}}\mu_{\mathrm{o}}-k_x^2\varepsilon_{\mathrm{o}}/\varepsilon_{\mathrm{e}}}.
    \label{eq: Dispersion TM}
\end{equation}
From eq.~\eqref{eq: Dispersion TE} and eq.~\eqref{eq: Dispersion TM}, it is apparent that anisotropy only plays a role for TM polarised light and is inherently dependent on the angle of incidence, as this determines the vector $k_x$ and thus the fraction of the electric field in the extraordinary direction. To determine our $k_x$ in the medium, we can use the most general version of Snell's law,
\begin{equation*}
    k_{1x}=k_{2x}.
\end{equation*}
Writing it in this way, avoids any ambiguity about which refractive index and refracted angle we refer to; for a birefringent material, such as the uniaxial medium considered here, this may not be obvious. The effect on the anisotropy on the reflectance of TM polarised light must be partially dependent on the thickness of the slab, due to the product $Tk_{2z}$.
Lastly, it should be noted that colours are calculated based on reflectance, $\Tilde{R}$, and not the reflection coefficient/reflectivity, $R$, but the two are related as $\Tilde{R}=\lvert R\rvert^2$.

\subsection{Reflectance of nanosphere arrays}
The scattering of light from a nanostructure can be described using a transition matrix (T-matrix), which links the incoming field to the outgoing field~\cite{Waterman1971ScatteringMatrix,MishchenkoLightBySmallParticles}. For a single spherical NP, this matrix contains the Mie coefficients~\cite{Mie1908} which are dependent on the geometry and material of the particle and can also be used to calculate the optical cross sections of the particles. This is how the optical response is usually quantified. For an array of spherical NPs, a similar approach can be utilised; here, the optical response is defined not just by the T-matrix of the individual particle, but by a sum over all of the interactions between the particles in the lattice. This lattice sum gives rise to lattice resonances\cite{Alvarez-Serrano2024BipartiteArrays,Manjavacas2019NearFieldNanoArray}, which have the potential to affect the overall response of the particle. Based on this summation, the scattering matrix (S-matrix) can be calculated, which is used to determine reflection and transmission of the array. In this paper we employ the software package Treams~\cite{Beutel2024Treams} to calculate the T-~and S-matrices and the spectra of the investigated arrays. For all calculations of T-matrices for the spherical NPs, the calculations have been performed including terms up to $\ell=5$ in the spherical harmonic vector functions. For the calculations of the lattice interaction with a plane wave, the parameter $b$, which defines a distance in reciprocal space, has been set as 0.01 with increments of 0.01 until convergence is obtained.

\subsection{Colourimetry}
Reflectance spectra are useful in describing light--matter interactions, but when evaluating structural colours the interest lies in how the colour will be perceived by the human eye. This is dependent on the spectral irradiance of light source in the visible region of the electromagnetic spectrum as well as the sensitivity of the cones in our eyes towards different colours, as described by the colour matching functions. To compute the resulting colours, standard equations for calculating tristimulus  values XYZ \cite{MalacaraColorimetry} have been used with the colour-matching functions from 1931 with a 2-degree observer\cite{CIE2019ColourMatch1931} and the D65 illuminant as light source \cite{CIE2019IlluminantD65}. Conversion of these tristimulus values to color coordinates in the standard RGB and CIELAB color spaces \cite{MalacaraColorimetry} was carried out to easily showcase the reflected color in conventional display devices and to quantify perceptual differences between colors, respectively. In the latter case, we have considered the parameter $\Delta E_{Lab} \equiv \Delta E_{00}$, defined according to the CIEDE 2000 convention of the International Commission on Illumination \cite{Sharma2005CIED2000ColorDiff,Luo2001DevelopmentCIED2000}, to characterize the difference between two colors. Here, the size of $\Delta E_{Lab}$ determines how easily a difference in colour is perceived, ranging from no perceived difference of any observer at $0<\Delta E_{Lab}<1$ and an observer will see two completely distinct colours at $\Delta E_{Lab}>5$\cite{Mokrzycki2011DeltaE}; there is further delimitation in what the size means which will be introduced if necessary. 

To infer on the gamut of reflected colors, we have utilised the CIEXYZ colourspace as defined by the International Commission on Illumination (CIE), where the space is presented as a horseshoe-like shape, where the values at the locus correspond to every pure monochromatic hues and the area delimited by it contains all physically realisable colours. The fraction of this colourspace which is reproducible by most digital displays -- the RGB colour space -- is also indicated on the diagrams presented in this work by a triangle with black sides.

\section{Results and discussion} 
\label{sec:results_discussion}
The investigation of TMD nanostructures as a platform for structural colour is multifaceted. Before the investigation of arrays from spherical NPs can take place, we would like to explore TMDs as a platform more thoroughly -- to that end, we explore the contributions of individual mechanisms, and in particular how the presence of excitons and the intrinsic TMD anisotropy affect structural colours.

\subsection{Excitonic effects}
Why choose TMDs as opposed to other HIDs? HIDs are useful in many optical applications; for structural colours particularly so, due to their ability to sustain Mie modes\cite{Mie1908, todisco2020EPolaritonsSiDisk, Verre2019TMDMieNanodisc}. While it has already been shown that HID nanostructures can sustain structural colours\cite{Daqiqeh2021NanoStructuralColours,Kristensen2016PlasmonColourGen,Cao2010SiliconWireColour, Faluraud2017SiliconFullColourPrints,Yang2020StructuralColoursSi3N4, Yang2020DielectricHighPerformance}, the addition of excitonic transitions of TMDs might lead to a larger tunability of the reflectance spectra. 

\begin{figure*}[ht]
    \centering
    \includegraphics[width=0.88\linewidth]{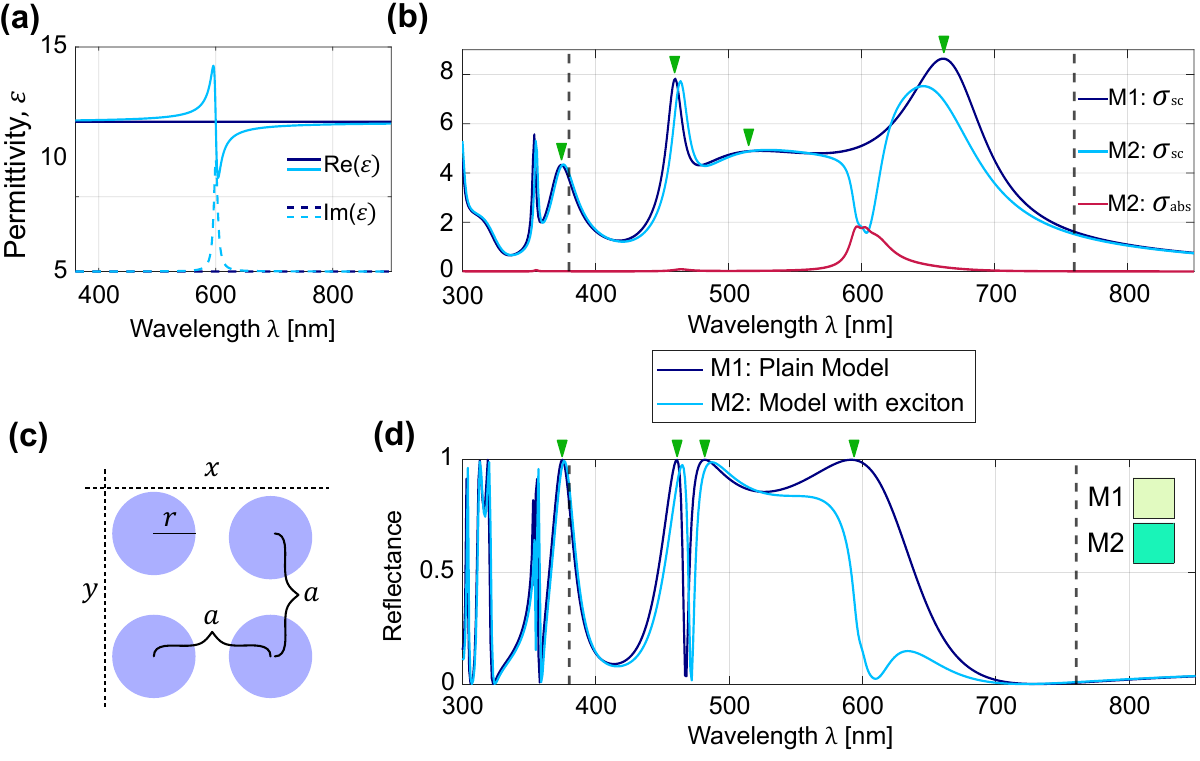}
    \caption{Illustration of the differences between a model of a lossless HID compared to a toy model, with the same base permittivity with one Lorentzian added to represent an exciton at 600 nm. Panel (a) shows the real (full lines) and imaginary part (dashed lines) of the permittivities of the plain model (M1, dark blue line) and the toy model (M2, light blue line). Panel (b) shows the normalised scattering cross sections, $\sigma_{\mathrm{sc}}$, of a single NP of the array for both the plain model (M1, dark blue line) and the toy model (M2, light blue line) as well as the normalised absorption cross section, $\sigma_\mathrm{abs}$, of the toy model (M2, red line). The plain model exhibits no absorption and its $\sigma_\mathrm{abs}$ is thus left out. A sketch of the unit cell is shown in panel (c). Panel (d) shows the reflectance spectra of arrays of spherical particles of either material; for both arrays the radius, $r$, of the spherical particles is 100~nm and the lattice constant, $a$, is 300~nm. Within the panel the colours produced by both arrays are shown in small squares.}
    \label{fig:Excitonic}
\end{figure*}
\begin{figure*}[ht!]
    \centering
    \includegraphics[width=0.99\linewidth]{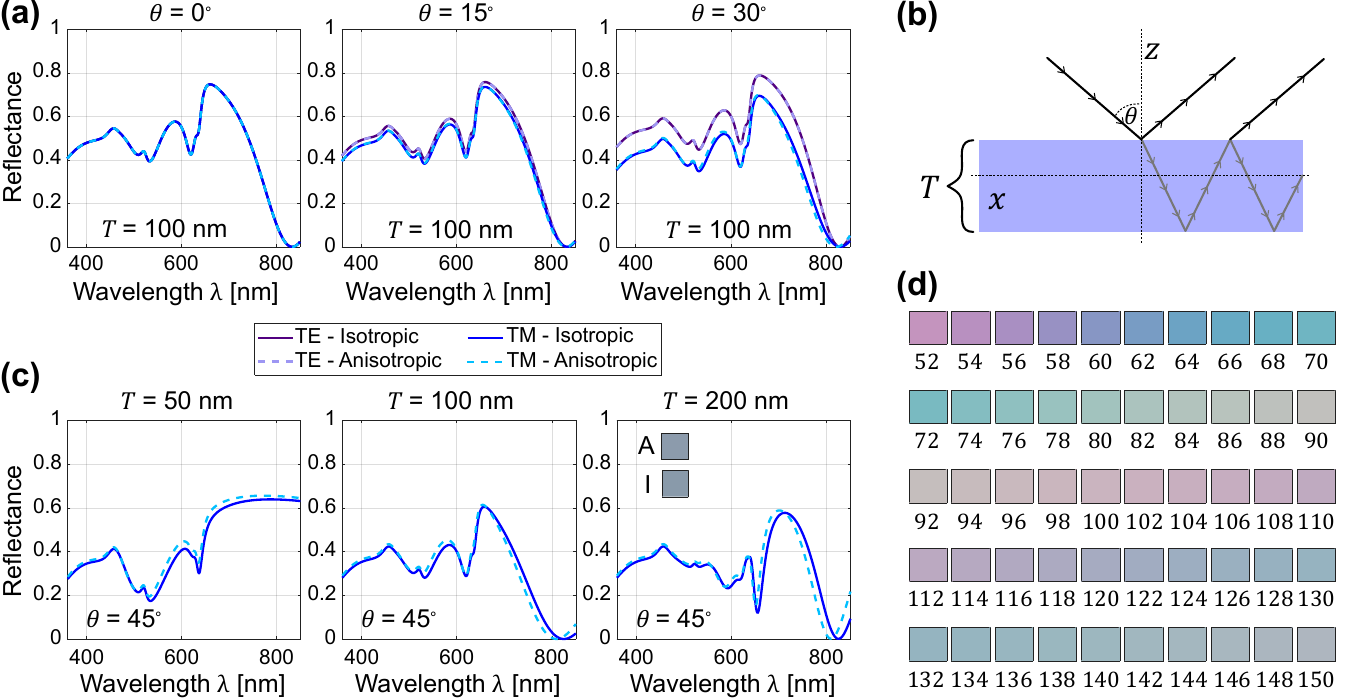}
    \caption{Visualisations of the investigation of a WS$_2$ slab. (a) Reflectance of TE and TM light impringing on a $T = 100$~nm slab, illustrated by the sketch in (b), when assumed isotropic or anisotropic for three incidence angles, $\theta$. (c) Reflectance of TM light at $\theta=45^\circ$ for varying thicknesses when the slab is assumed isotropic or anisotropic. (d) Colours stemming from the reflectance spectra for isotropic WS$_2$ slabs, with thicknesses varying from 52~nm to 150~nm in steps of 2~nm, as indicated  by the numbers below each square.}
    \label{fig:BigAnisotropy graph}
\end{figure*}
Consider Figure~\ref{fig:Excitonic} where panel (a) shows the permittivity of a basic HID $\varepsilon_1 = 10$ as M1, and a toy model where a single exciton is added to this basic model at 600~nm as M2. The permittivity of M2 is calculated as $\varepsilon_2 = \varepsilon_1 + f_k\lambda/(\lambda_k^2-\lambda^2-\lambda\gamma_k\mathrm{i})$, where $\lambda$ is the wavelength, $f_k=60$~nm is the oscillator strength, $\gamma_k=8$~nm is the damping/linewidth and $\lambda_k=600$~nm is the central wavelength of the excitonic transition.
These two permittivities are used as the material properties of a single spherical NP of radius $r=100$~nm, which produces the Mie-dominated scattering cross sections in Figure~\ref{fig:Excitonic}(b). The peaks in scattering come from Mie resonances, which arise as the electric field induces local pairs of polarisation charges throughout the particle that oscillate in phase with the electric field.\cite{HohenesterNanoQuantumOptics}. For small particles these oscillations correspond to an effective electric dipole; as the size of the particle increases higher order multipoles can be sustained, and retardation causes a displacement current which induces magnetic multipoles. Taking a closer look at the spectrum of M1 in Figure~\ref{fig:Excitonic}(b), we consider the peaks marked by green pointers in descending order: the large peak at around 660~nm represents the magnetic dipole contribution; the electric dipole contribution is quite wide but its top can be glanced as the slight bump at around 515~nm; the sharper peaks at 460~nm and 375~nm are, in order, the magnetic and electric quadrupolar contributions and below these wavelengths further higher-order multipoles emerge. If NPs of this kind are arranged in some structure, their interaction will shift and morph the resulting spectrum, but it should still be expected that these resonances are identifiable.
The NPs are here placed in an infinite square array with lattice constant $a=300$~nm, as sketched in Figure~\ref{fig:Excitonic}(c). The array lies in the $xy$-plane and is excited normally with light which is polarised in the $y$-direction, and the resulting reflectance spectrum is calculated for wavelengths in the visible range
(and slightly beyond). All of this is done using the software package Treams~\cite{Beutel2024Treams} (see Methods). Considering Figure~\ref{fig:Excitonic}(d) in tandem with our analysis of the single particle spectrum, some of the spectral features can easily be identified as stemming from specific resonances; these features are again marked by green pointers. The large band of reflectance with soft peaks at around 480~nm and 590~nm (for the model without an exciton) must stem from the electric and magnetic dipole respectively. Moving to the left, there are two sharp peaks which are placed at 460~nm and 375~nm exactly like the quadrupolar response of the single particle, making it clear that for a simple permittivity the reflectance spectrum is closely connected to the scattering cross section.\\
As can also be seen in Figure~\ref{fig:Excitonic}(d), an exciton at 600~nm heavily suppresses the reflectance, which does cause a significant shift in the resulting colour [the colours for each model can be seen in the inset in (d)] with a $\Delta E_{Lab}=19.4$. One may be tempted to assume that this effect is a splitting of modes due to hybridisation between the geometric modes and the exciton; however, our inspection of the absorption spectrum shows that it is primarily due to the exciton absorbing the incident light at this wavelength -- this can also be seen by looking at the scattering and absorption cross sections from a single spherical particle in Figure~\ref{fig:Excitonic}(b). If strong hybridisation took place, further differentiation in the colours could possibly be seen. From this it can be expected that excitons may play an interesting role in the colour-possibilities stemming from HID nanostructures.

\subsection{Anisotropy in TMD slabs}
To investigate the effect of anisotropy on the reflectance spectrum and thereby determine if the full tensorial expressions of the permittivity and permeability are needed, slabs of TMDs are investigated. For all results shown in this paper, ordinary and extraordinary permittivity of WS$_2$  obtained from experimental data by Vyshnevyy et al.~\cite{Vyshnevyy2023WS2} have been used. As mentioned in the methodology section, TMDs are anisotropic and thus obey slightly different equations for the reflection coefficients, which affects the TM-polarised light and by extension may affect the colours produced. In this section, we show only viewing angles up to $45^\circ$ having assessed this as most pertinent, however, an extended analysis to $75^\circ$ can be seen in Figure~S.1 in the Supporting Information (SI), resulting in a near-identical conclusion as in this section.
\begin{figure*}[ht!]
    \centering
    \includegraphics[width=1\linewidth]{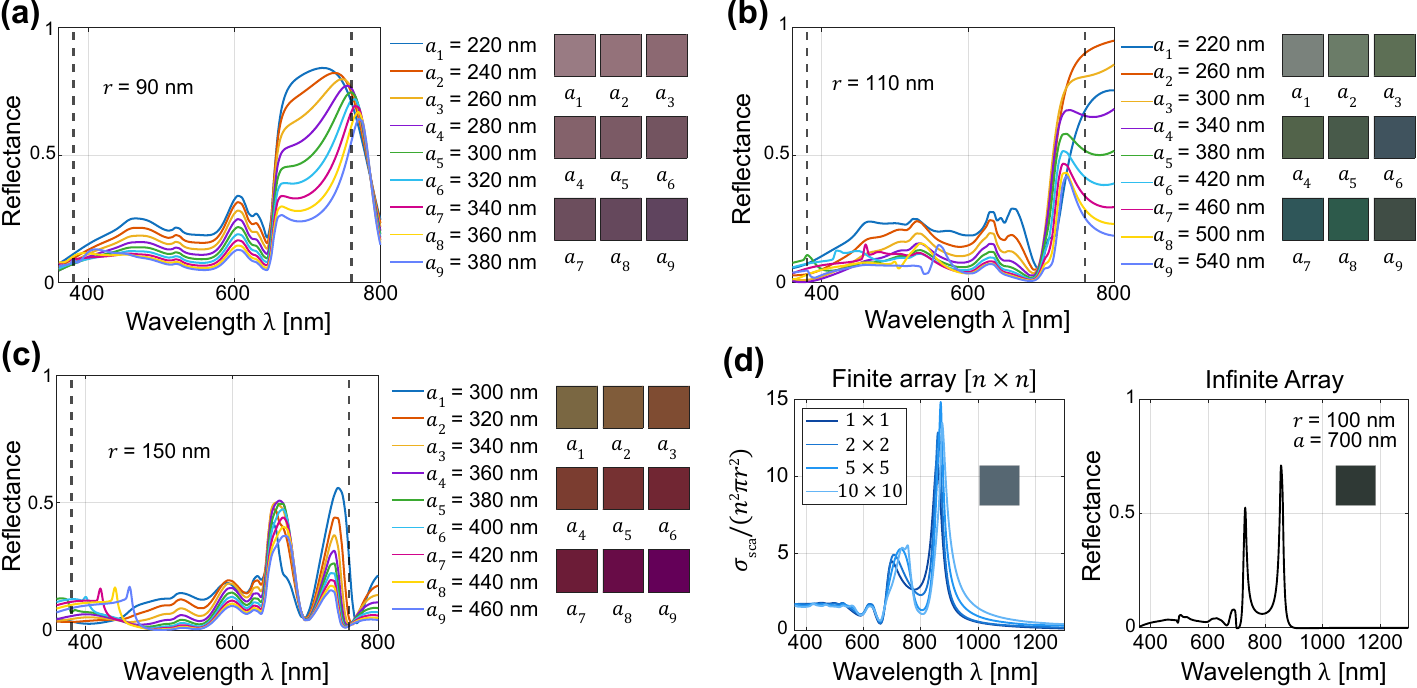}
    \caption{Visualisations of data obtained from arrays of spherical NPs. Panels (a), (b), and (c) show reflectance spectra and corresponding colours for infinite arrays of spherical NPs, with variable lattice constants and radii of (a) 90~nm, (b) 110~nm, and (c) 150~nm. Panel (d) shows the normalised scattering cross section of finite arrays of varying sizes and the reflectance spectra of an infinite array of spherical WS$_2$ NPs, with radius $r=100$~nm and lattice constant $a=700$~nm. The inset in each subpanel shows the colours of a single NP and the colour of the infinite array, respectively.}
    \label{fig:Array_colours}
\end{figure*}
\begin{figure*}[ht]
    \centering
    \includegraphics[width=0.99\linewidth]{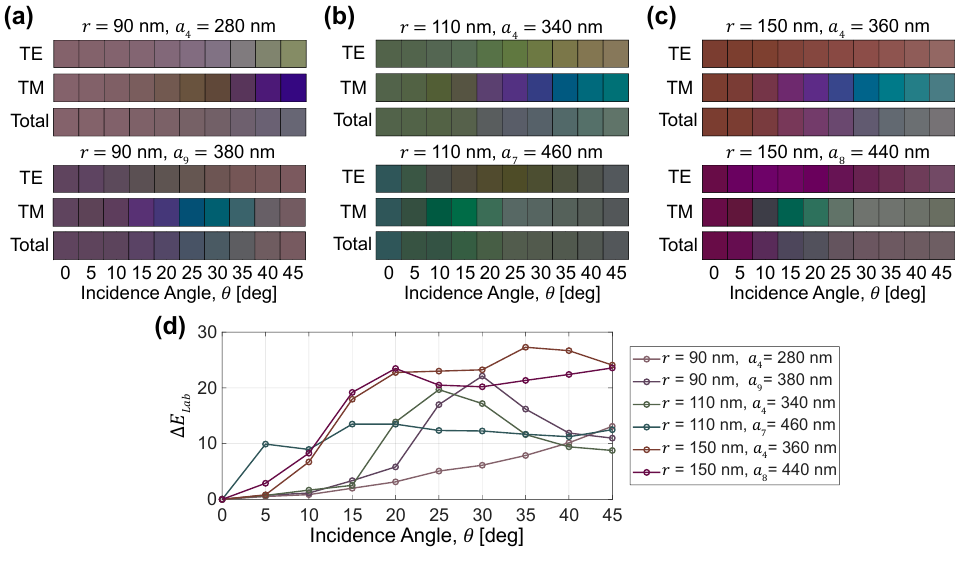}
    \caption{Colours obtained from selected arrays with NPs of radius (a) 90 nm (b) 110 nm and (c) 150 nm with a variation of incidence angle. Each panel contains the colours from the TE, TM and total (average of TM and TE) reflectances as the angle of incidence is varied from $0^\circ$ to $45^\circ$. (d) The colour difference, $\Delta E_{Lab}$, between the average colour at each incidence angle with respect to the colour at normal incidence.}
    \label{fig:Array_angles}
\end{figure*}
As can be seen in Figure~\ref{fig:BigAnisotropy graph}(a), at normal incidence anisotropy is irrelevant even for TM polarised light, and the reflectances of the two polarisations are identical. As the incidence angle is increased, this is no longer the case, as can be seen in the next two panels in  Figure~\ref{fig:BigAnisotropy graph}(a). First of all, when the angle is increased the reflections of the TE and TM polarisations begin to differ. From a quick inspection it is also clear that the features of the spectra are shifted between the isotropic and anisotropic case for the TM polarisation. Upon numerical inspection of the largest global minima and maxima, we find that there is a negligible change in the value of the reflectance for the main resonant features (on the 2nd or 3rd decimal), but some amount of blueshift for the anisotropic case compared to the isotropic. The largest shifts occur for the largest incidence angle, as can be seen when comparing Figure~\ref{fig:BigAnisotropy graph}(a) ($\theta=15^\circ$ and $\theta=30^\circ$) to the second panel in Figure~\ref{fig:BigAnisotropy graph}(c) ($T=100$~nm, $\theta=45^\circ$). For the latter slab, the peaks just before and after 600~nm are shifted by around 4~nm, but the biggest shift occurs at the minimum by 16~nm. This inadvertently shows the wavelength dependence of the anisotropic effect coming from the product $Tk_{2z}$ in eq.~\eqref{eq: Reflectance_Slab}. The effect of these shifts we would expect, however, to be relatively small in the context of colour analysis.

Combining the effect of both the angle and thickness, as shown in  Figure~\ref{fig:BigAnisotropy graph}(c), the anisotropy does seem to become more prominent for increasing thicknesses. However, even for $T=200$~nm the largest shift of a maximum is of 11~nm and the shift of the primary minimum is still 16~nm. 
When determining the colour based on the two different reflectances the change is almost imperceptible as indicated by a $\Delta E_{Lab}=1.19$, where only a trained observer may notice a difference. It should further be considered that this effect is only for half of the total reflectance of unpolarised light (with TE polarisation comprising the other half), in which case the difference should become even more negligible. When performing the same analysis for MoS$_2$ (using experimental data from Ermolaev et al.~\cite{Ermolaev2020MoS2}), the effects are even less intense. Here, the largest shifts are 10~nm and most features are shifted by less than 5~nm. Similar analysis for HfSe$_2$ (using ab initio calculations from Zotev et al.~\cite{Zotev2023VanDerWaals}) shows no shifts in wavelength for the prominent features and only small, nearly negligible changes in intensity. We thus conclude that for the purposes of this paper, where only the visible spectrum is concerned, variations in angular incidence are far below 90$^\circ$ and material thicknesses do not exceed 200~nm, the anisotropy of the TMDs is negligible and they can be assumed isotropic using the ordinary index.\\
As can be seen in Figure~\ref{fig:BigAnisotropy graph}(d), slabs can produce colours due to the Fabry-P\'{e}rot cavity resonances stemming from the interference of all the reflected rays\cite{Daqiqeh2021NanoStructuralColours}, however, as the slab gets thicker the colour tends towards the metallic grey which is commonly associated with the bulk. In addition, the tailorability of the colours is very limited as the only tuning parameter is the thickness of the slab. Consequently, moving to more complex nanostructures, as presented in the next section, may allow to broaden the gamut of achievable colours without changing the material.

\subsection{Arrays of NPs}
To increase tuneability, we move from slabs to arrays of spherical WS$_2$ NPs; here, there are several tuning parameters such as the radius of the particles or the lattice constants in either directions.
This investigation considers infinite square arrays ($a_x=a_y$) with particles of identical radius, unless otherwise stated. The permittivity is once again taken from experimental results\cite{Vyshnevyy2023WS2}; 
based on the analysis in the previous section the NPs are treated as isotropic. To gain insight in a large combination of radius and lattice constants a sweep was undertaken with radius varying from 30~nm to 150~nm and lattice constants ranging from the diameter of the particles (meaning that the adjacent NPs would just touch) to 800~nm.
These sweeps have generated more spectra and colours than is reasonable to show in this work, however, a small selection is shown in Figure~\ref{fig:Array_colours}(a)(b)(c), where each subfigure represents arrays with particles of the same radius but each line and colour has a different associated lattice constant. As can be seen, the arrays are able to create a variety of colours -- for some radii the range of colours generated is larger than others, but overall it is clear that both the lattice constant and the radius play a role. The fact that the radius has an effect on the resulting colour may be the most obvious, as the primary optical response of dielectric NPs come from Mie resonances which are highly dependent on the size of the NP. Several of the spectra also exhibit some amount of self-hybridisation around the 620~nm exciton, suggesting that for TMDs with excitons of higher oscillator strength this may present itself as a viable mechanism for tuning the spectral reflectance. On the other hand, lattice resonances did not show themselves to be a very relevant mechanism to tune the optical response and by extension the colours of the arrays. For many of the arrays in Figure~\ref{fig:Array_colours}(a)(b)(c) the lattice resonances do not even appear, as the lattice constant is smaller than 360~nm which is outside the range of the plot; for the ones where they are visible they are primarily very small bumps in reflectance (most easily seen for $a_6$ through $a_9$ in Figure~\ref{fig:Array_colours}(c)). In these graphs only the reflection is presented; it should be noted that for most of the arrays the transmission was relatively high where the reflectance is low (see SI, Figure~S.2), thus the arrays may appear slightly glassy.
\begin{figure*}[ht]
    \centering
    \includegraphics[width=0.65\linewidth]{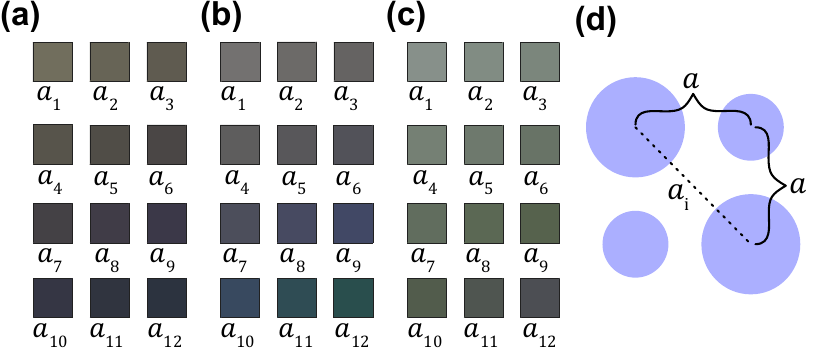}
    \caption{Colours obtained from arrays with NPs of radius (a) 70~nm, (b) 100~nm and (c) the combined bipartite array at the same lattice constants varying from $a_1=260$~nm to $a_{12}=480$~nm in steps of 20~nm. An illustration of the configuration of particles in the bipartite array can be seen in (d).}
    \label{fig:bipartite}
\end{figure*}
As mentioned in the section on methodology, the S-matrix of an array is dependent on the optical response of the NPs and lattice sums, and while semi-analytical methods allow the response of an array to be calculated reasonably quickly, the analytical scattering cross section calculation of the single particle will always be faster. Thus, if the scattering cross section of a single particle was a good representative of lineshape of the reflection spectrum, with only a lattice resonance missing (as one might anticipate in the case of sparse arrays), this would be an excellent method for estimating expected colours from an array at a lower computational cost.
In Figure~\ref{fig:Array_colours}(d), the spectra of a single particle and an infinite array are shown to illustrate the connection between them and to investigate the importance of the lattice compared to the single NP optical response. Here we show the normalised scattering cross section of a single particle, but also how the scattering cross-section changes as additional particles are added, making small finite square arrays. Next to it, we show the reflectance spectrum of the infinite array, with particles of the same radius and same lattice constant as the finite arrays. As the number of particles in the array increases, the spectral features are sharpened, and the lattice mode appears to split the secondary maxima (see dip at around 700~nm). While the two spectra share features, there are still significant shifts between them so that the hue of the array cannot be predicted by the single particle. This is further illustrated by the insets which show the colour of the single NP as estimated from the scattering cross section (after it has been normalised to lie between 0 and 1) and the infinite array; these two colours are significantly different, as indicated by their $\Delta E_{Lab}=17.8$. Furthermore, the exact lightness of the colour can never be precisely predicted from the scattering, as the intensity cannot clearly be translated to reflectance.

Being dependent on geometry means that structural colours are often also dependent on the viewing angle, thus for our evaluation of the structures this must also be taken into account. Two arrays from each of the panels in Figure~\ref{fig:Array_colours}(b)(c)(d) have been chosen for further investigation; here the incidence angle is increased from $0^\circ$ to $45^\circ$ in steps of $5^\circ$ and the arrays are subjected to both TE and TM polarised light. Based on the analysis made in the previous section the particles can be treated as isotropic for these angles and thicknesses. The resulting reflectance spectra, as well as the average reflectance spectra, $\frac{\Tilde{R}^{\mathrm{(TE)}}+\Tilde{R}^{\mathrm{(TM)}}}{2}$ (intended to represent true unpolarised light), are then analysed to determine the change in reflected colour with the change in angle of incidence.

The results of this treatment are shown in Figure~\ref{fig:Array_angles}, where the progression of colours can be seen as the angle is increased. For all arrays, it is very apparent how TE and TM polarised light have differing reflectance spectra, with the colour generally varying more for the TM polarisation. For all of the arrays, it is also clear that there will be some angular variance in the perceived colour from unpolarised light; here the arrays in (b) and the top panel in (a) fare the best throughout the angular variation, which can also be seen by examining the trends in $\Delta E$ as seen in Figure~\ref{fig:Array_angles}(d). It may also be interesting to note that, for the colours in panel (c) in particular, the TE polarised light has a much more consistent colour with angular variation, suggesting that if combined with a polariser in some way it can lead to angularly invariant colours. Having this colour variance is a drawback; however, it is one which is common throughout structural colours~\cite{Daqiqeh2021NanoStructuralColours} and may be used as a feature in some use-cases such as having a decorative purpose in consumer products.

The arrays investigated so far are only a small fraction of possible configurations of arrays -- one extension may be utilising bipartite arrays where particles of two different radii alternate in both the $x$- and $y$-direction. A unit cell of such an array can be seen in Figure~\ref{fig:bipartite}(d). For better comparison between the earlier arrays and the bipartite arrays, the overall lattice constant of the new arrays will be considered as the distance between adjacent particles of the same radius, denoted as $a_i$ on the sketch.

In Figure~\ref{fig:bipartite}(a) and (b) the colours of the individual arrays and in (c) their combined bipartite array can be seen. This illustrates that the combination of arrays can lead to colours which are not represented by either individual array, suggesting that further combinations might lead to extending the coverage of the colour space. The reflectances of the arrays can be seen in SI Figure~S.3.

To show the coverage of the colour space from the investigated arrays at normal incidence we consider the CIEXYZ diagrams, where the locus of the shape corresponds to pure monochromatic colours and the coloured space corresponds to the entire visible colour space, so even beyond the RGB rendering capacity which is marked by the black triangle (see Methods for further description). On the plots below, each white dot represents one array.
\begin{figure*}[ht]
    \centering
    \includegraphics[width=0.99\linewidth]{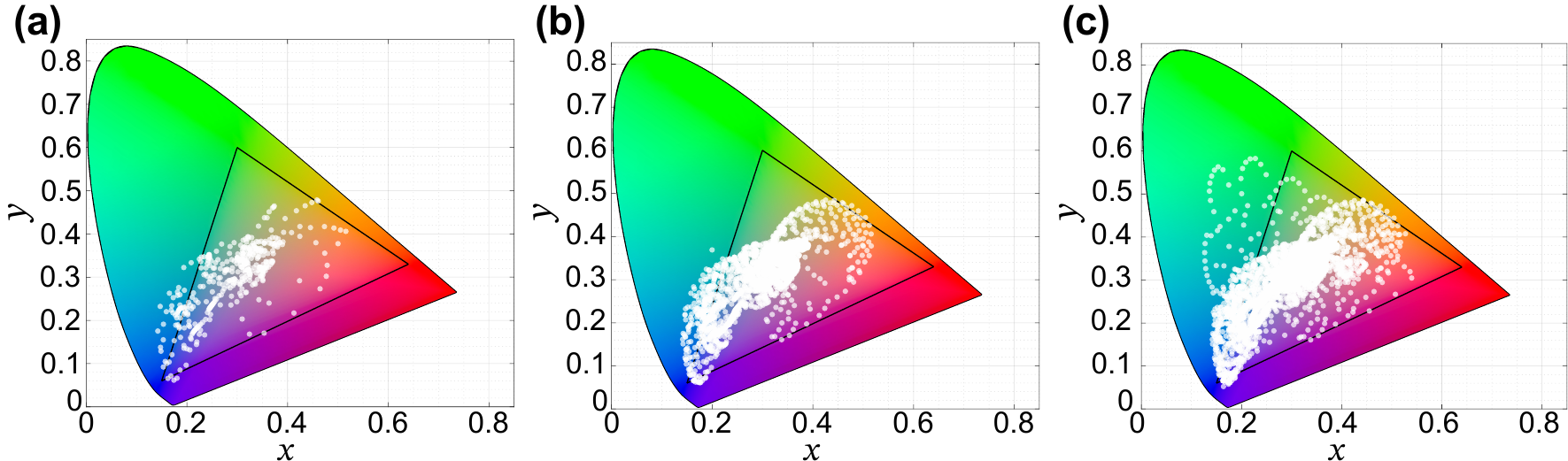}
    \caption{Plots of the CIE colour space, where each white dot represents a colour obtained by one of the arrays investigated. Subfigure (a) show arrays consisting of spherical WS$_2$ NPs, where the radius of the particles vary from 30~nm to 150~nm in steps of 10~nm, while the lattice constant varies from the diameter of the particles to 800~nm in steps of 20~nm. In (b) arrays with intermediate radii in steps of two have been added as well as selected bipartite arrays. In subfigure (c) arrays of the same structure as the arrays in (a) but utilising several different TMDs, here MoS$_2$\cite{Ermolaev2020MoS2}, MoSe$_2$\cite{Zotev2023VanDerWaals}, HfS$_2$\cite{Zhao2017DielectricProperties} and HfSe$_2$\cite{Zotev2023VanDerWaals}.}
    \label{fig:CIE_spheres}
\end{figure*}
In Figure~\ref{fig:CIE_spheres}(a) square arrays of WS$_2$ NPs of radii varying from 30 to 150~nm in 10~nm steps and lattice constants varying from the diameter of the particles to 800~nm in steps of 20~nm are included. Here a little under half the RGB triangle is covered, with especially the red and green corners lacking. However, the areas of the colour space which are partially filled could be filled more densely by including smaller steps in both radius and periodicity and including combinations of bipartite arrays. This can be seen in Figure~\ref{fig:CIE_spheres}(b) where the radii for the arrays now vary with just 2~nm between each array (though the lattice constants still vary with 20~nm and have been rounded to the nearest 10 at the beginning to be more comparable) as well as including bipartite arrays with 6 different radii configurations. To reach into further corners of the colour spectrum, expanding the selection of materials has been considered. In Figure~\ref{fig:CIE_spheres}(c) arrays of particles made from HfS$_2$, HfSe$_2$, MoS$_2$ and MoSe$_2$ have been added. These have the same dimensionality as the initial WS$_2$ arrays. Now most of the RGB space is reached, and once again it will be more densely filled if additional arrays with in-between radii and lattice constants are added. Further coverage of the colour space may also be obtained through investigations of different overarching structures, for example: changing the type of lattice to a rectangular or even hexagonal one; testing out the layering of different arrays; or even testing the effect of different substrates.

In recent years methods for synthesising TMD nanostructures and NPs either colloidally or utilising self-assembly have been achieved\cite{Li2024ColloidalSynethesisTMD,Sun2021ColloidalStructure,Bisen2022SelfAssemblyTMDs}, with some even achieving particles with a size range of a few to hundreds of~nm and spherical enough to exhibit theoretically predicted Mie resonances\cite{Gleb2022TMDNanoSpheres}.
Combined with methods like optical force stamping lithography\cite{Nedev2011OFSL} it may very well be possible to achieve the nanostructures considered here. Though for different fabrication capabilities, the analysis here may also be valid for nanodiscs of comparable dimensions. Nanodiscs where the height and diameter are not equal might also present further interesting possibilities within structural colours, as the symmetry breaking causes shifts between the electric and magnetic scattering Mie modes of each particle\cite{Staude2013TailoringScattering}, potentially leading to even further tuneability in the colour range\cite{Verre2019TMDMieNanodisc}.
\section{Conclusion}
 In this work, we presented TMD nanostructures as a possible platform for structural colours due to both the HID features of these materials and the excitons that they sustain. These excitons can directly affect the reflectance spectrum either through absorption or by hybridising with the geometric modes of the NPs. We have also shown that while TMDs such as WS$_2$ are undeniably anisotropic, the effect of this anisotropy on the structural colours is so negligible that we may ignore it for ease of calculation. However, as with many structures, the viewing angle does affect the spectrum and thus the colour which should be considered if these types of structures are to be implemented. Most importantly, we showed that by varying few parameters with relatively large step-sizes it is possible to cover most of the RGB colour space, and proposed that most colours can be realised if more combinations are considered.
\section*{Acknowledgments}
We acknowledge support from the Independent Research Fund Denmark (Grant No. 5281-00155B). The Center for Polariton-driven Light--Matter Interactions (POLIMA) is funded by the Danish National Research Foundation (Project No. DNRF165, POLIMA).
We thank Joel D. Cox for proofreading and commenting on the manuscript.
\section*{Author contribution}
C.~T. conceived the idea. Y.~L. provided valuable insight on TMDs for material selection. C.~G.~F. provided codes and expertise within colourimetric calculations and analysis. I.~J.~B. performed all calculations. I.~J.~B. and C.~T. analysed the results.  I.~J.~B. wrote the original draft. All authors participated in editing and revising the final manuscript.

\hfill

\bibliography{references}

\begin{thebibliography}{65}%
\makeatletter
\providecommand \@ifxundefined [1]{%
 \@ifx{#1\undefined}
}%
\providecommand \@ifnum [1]{%
 \ifnum #1\expandafter \@firstoftwo
 \else \expandafter \@secondoftwo
 \fi
}%
\providecommand \@ifx [1]{%
 \ifx #1\expandafter \@firstoftwo
 \else \expandafter \@secondoftwo
 \fi
}%
\providecommand \natexlab [1]{#1}%
\providecommand \enquote  [1]{``#1''}%
\providecommand \bibnamefont  [1]{#1}%
\providecommand \bibfnamefont [1]{#1}%
\providecommand \citenamefont [1]{#1}%
\providecommand \href@noop [0]{\@secondoftwo}%
\providecommand \href [0]{\begingroup \@sanitize@url \@href}%
\providecommand \@href[1]{\@@startlink{#1}\@@href}%
\providecommand \@@href[1]{\endgroup#1\@@endlink}%
\providecommand \@sanitize@url [0]{\catcode `\\12\catcode `\$12\catcode `\&12\catcode `\#12\catcode `\^12\catcode `\_12\catcode `\%12\relax}%
\providecommand \@@startlink[1]{}%
\providecommand \@@endlink[0]{}%
\providecommand \url  [0]{\begingroup\@sanitize@url \@url }%
\providecommand \@url [1]{\endgroup\@href {#1}{\urlprefix }}%
\providecommand \urlprefix  [0]{URL }%
\providecommand \Eprint [0]{\href }%
\providecommand \doibase [0]{https://doi.org/}%
\providecommand \selectlanguage [0]{\@gobble}%
\providecommand \bibinfo  [0]{\@secondoftwo}%
\providecommand \bibfield  [0]{\@secondoftwo}%
\providecommand \translation [1]{[#1]}%
\providecommand \BibitemOpen [0]{}%
\providecommand \bibitemStop [0]{}%
\providecommand \bibitemNoStop [0]{.\EOS\space}%
\providecommand \EOS [0]{\spacefactor3000\relax}%
\providecommand \BibitemShut  [1]{\csname bibitem#1\endcsname}%
\let\auto@bib@innerbib\@empty
\bibitem [{\citenamefont {Rezaei}\ \emph {et~al.}(2021)\citenamefont {Rezaei}, \citenamefont {Dong}, \citenamefont {Chan}, \citenamefont {Trisno}, \citenamefont {Ng}, \citenamefont {Ruan}, \citenamefont {Qiu}, \citenamefont {Mortensen},\ and\ \citenamefont {Yang}}]{Daqiqeh2021NanoStructuralColours}%
  \BibitemOpen
  \bibfield  {author} {\bibinfo {author} {\bibfnamefont {S.~D.}\ \bibnamefont {Rezaei}}, \bibinfo {author} {\bibfnamefont {Z.}~\bibnamefont {Dong}}, \bibinfo {author} {\bibfnamefont {J.~Y.~E.}\ \bibnamefont {Chan}}, \bibinfo {author} {\bibfnamefont {J.}~\bibnamefont {Trisno}}, \bibinfo {author} {\bibfnamefont {R.~J.~H.}\ \bibnamefont {Ng}}, \bibinfo {author} {\bibfnamefont {Q.}~\bibnamefont {Ruan}}, \bibinfo {author} {\bibfnamefont {C.-W.}\ \bibnamefont {Qiu}}, \bibinfo {author} {\bibfnamefont {N.~A.}\ \bibnamefont {Mortensen}},\ and\ \bibinfo {author} {\bibfnamefont {J.~K.~W.}\ \bibnamefont {Yang}},\ }\href {https://doi.org/10.1021/acsphotonics.0c00947} {\bibfield  {journal} {\bibinfo  {journal} {ACS {P}hotonics}\ }\textbf {\bibinfo {volume} {8}},\ \bibinfo {pages} {18} (\bibinfo {year} {2021})}\BibitemShut {NoStop}%
\bibitem [{\citenamefont {Kinoshita}\ \emph {et~al.}(2002)\citenamefont {Kinoshita}, \citenamefont {Yoshioka},\ and\ \citenamefont {Kawagoe}}]{Kinoshite2002MorphoButterfly}%
  \BibitemOpen
  \bibfield  {author} {\bibinfo {author} {\bibfnamefont {S.}~\bibnamefont {Kinoshita}}, \bibinfo {author} {\bibfnamefont {S.}~\bibnamefont {Yoshioka}},\ and\ \bibinfo {author} {\bibfnamefont {K.}~\bibnamefont {Kawagoe}},\ }\href {https://doi.org/10.1098/rspb.2002.2019} {\bibfield  {journal} {\bibinfo  {journal} {Proceedings of the Royal Society B: Biological Sciences}\ }\textbf {\bibinfo {volume} {269}},\ \bibinfo {pages} {1417} (\bibinfo {year} {2002})}\BibitemShut {NoStop}%
\bibitem [{\citenamefont {Clausen}\ \emph {et~al.}(2014)\citenamefont {Clausen}, \citenamefont {Højlund-Nielsen}, \citenamefont {Christiansen}, \citenamefont {Yazdi}, \citenamefont {Grajower}, \citenamefont {Taha}, \citenamefont {Levy}, \citenamefont {Kristensen},\ and\ \citenamefont {Mortensen}}]{Clausen2014ColourationPlastic}%
  \BibitemOpen
  \bibfield  {author} {\bibinfo {author} {\bibfnamefont {J.~S.}\ \bibnamefont {Clausen}}, \bibinfo {author} {\bibfnamefont {E.}~\bibnamefont {Højlund-Nielsen}}, \bibinfo {author} {\bibfnamefont {A.~B.}\ \bibnamefont {Christiansen}}, \bibinfo {author} {\bibfnamefont {S.}~\bibnamefont {Yazdi}}, \bibinfo {author} {\bibfnamefont {M.}~\bibnamefont {Grajower}}, \bibinfo {author} {\bibfnamefont {H.}~\bibnamefont {Taha}}, \bibinfo {author} {\bibfnamefont {U.}~\bibnamefont {Levy}}, \bibinfo {author} {\bibfnamefont {A.}~\bibnamefont {Kristensen}},\ and\ \bibinfo {author} {\bibfnamefont {N.~A.}\ \bibnamefont {Mortensen}},\ }\href {https://doi.org/10.1021/nl5014986} {\bibfield  {journal} {\bibinfo  {journal} {Nano {L}etters}\ }\textbf {\bibinfo {volume} {14}},\ \bibinfo {pages} {4499} (\bibinfo {year} {2014})}\BibitemShut {NoStop}%
\bibitem [{\citenamefont {Ferreira}\ \emph {et~al.}(2025)\citenamefont {Ferreira}, \citenamefont {Paul}, \citenamefont {Babin}, \citenamefont {Lamminaho}, \citenamefont {Andersen}, \citenamefont {Thorsteinsson}, \citenamefont {Poulsen}, \citenamefont {Petersons}, \citenamefont {Yde}, \citenamefont {Stensborg}, \citenamefont {Mortensen}, \citenamefont {Cox},\ and\ \citenamefont {Madsen}}]{Ferreira2025AngularColourIntegratedPV}%
  \BibitemOpen
  \bibfield  {author} {\bibinfo {author} {\bibfnamefont {C.~G.}\ \bibnamefont {Ferreira}}, \bibinfo {author} {\bibfnamefont {A.}~\bibnamefont {Paul}}, \bibinfo {author} {\bibfnamefont {M.}~\bibnamefont {Babin}}, \bibinfo {author} {\bibfnamefont {J.}~\bibnamefont {Lamminaho}}, \bibinfo {author} {\bibfnamefont {N.~L.}\ \bibnamefont {Andersen}}, \bibinfo {author} {\bibfnamefont {S.}~\bibnamefont {Thorsteinsson}}, \bibinfo {author} {\bibfnamefont {P.~B.}\ \bibnamefont {Poulsen}}, \bibinfo {author} {\bibfnamefont {K.}~\bibnamefont {Petersons}}, \bibinfo {author} {\bibfnamefont {L.}~\bibnamefont {Yde}}, \bibinfo {author} {\bibfnamefont {J.~F.}\ \bibnamefont {Stensborg}}, \bibinfo {author} {\bibfnamefont {N.~A.}\ \bibnamefont {Mortensen}}, \bibinfo {author} {\bibfnamefont {J.~D.}\ \bibnamefont {Cox}},\ and\ \bibinfo {author} {\bibfnamefont {M.}~\bibnamefont {Madsen}},\ }\href {https://doi.org/https://doi.org/10.1002/solr.202500674} {\bibfield  {journal} {\bibinfo  {journal} {Solar RRL}\ }\textbf {\bibinfo {volume}
  {9}},\ \bibinfo {pages} {e202500674} (\bibinfo {year} {2025})}\BibitemShut {NoStop}%
\bibitem [{\citenamefont {Ferreira}\ \emph {et~al.}(2026)\citenamefont {Ferreira}, \citenamefont {Lamminaho}, \citenamefont {Paul}, \citenamefont {Babin}, \citenamefont {Andersen}, \citenamefont {Thorsteinsson}, \citenamefont {Poulsen}, \citenamefont {Petersons}, \citenamefont {Yde}, \citenamefont {Stensborg}, \citenamefont {Mortensen}, \citenamefont {Cox},\ and\ \citenamefont {Madsen}}]{Ferreira2026ColourPVOptimization}%
  \BibitemOpen
  \bibfield  {author} {\bibinfo {author} {\bibfnamefont {C.~G.}\ \bibnamefont {Ferreira}}, \bibinfo {author} {\bibfnamefont {J.}~\bibnamefont {Lamminaho}}, \bibinfo {author} {\bibfnamefont {A.}~\bibnamefont {Paul}}, \bibinfo {author} {\bibfnamefont {P.}~\bibnamefont {Babin}}, \bibinfo {author} {\bibfnamefont {N.~L.}\ \bibnamefont {Andersen}}, \bibinfo {author} {\bibfnamefont {S.}~\bibnamefont {Thorsteinsson}}, \bibinfo {author} {\bibfnamefont {P.~B.}\ \bibnamefont {Poulsen}}, \bibinfo {author} {\bibfnamefont {K.}~\bibnamefont {Petersons}}, \bibinfo {author} {\bibfnamefont {L.}~\bibnamefont {Yde}}, \bibinfo {author} {\bibfnamefont {J.~F.}\ \bibnamefont {Stensborg}}, \bibinfo {author} {\bibfnamefont {N.~A.}\ \bibnamefont {Mortensen}}, \bibinfo {author} {\bibfnamefont {J.~D.}\ \bibnamefont {Cox}},\ and\ \bibinfo {author} {\bibfnamefont {M.}~\bibnamefont {Madsen}},\ }\href {https://doi.org/https://doi.org/10.1016/j.nanoen.2025.111659} {\bibfield  {journal} {\bibinfo  {journal} {Nano Energy}\ }\textbf {\bibinfo
  {volume} {148}},\ \bibinfo {pages} {111659} (\bibinfo {year} {2026})}\BibitemShut {NoStop}%
\bibitem [{\citenamefont {Zhang}\ \emph {et~al.}(2019)\citenamefont {Zhang}, \citenamefont {Chen}, \citenamefont {Hu}, \citenamefont {Xu}, \citenamefont {Wang}, \citenamefont {Qin}, \citenamefont {Cao}, \citenamefont {Guan}, \citenamefont {Miroshnichenko}, \citenamefont {Gu},\ and\ \citenamefont {Li}}]{Zhang2019ColourSolarCellsLID}%
  \BibitemOpen
  \bibfield  {author} {\bibinfo {author} {\bibfnamefont {Y.}~\bibnamefont {Zhang}}, \bibinfo {author} {\bibfnamefont {S.}~\bibnamefont {Chen}}, \bibinfo {author} {\bibfnamefont {D.}~\bibnamefont {Hu}}, \bibinfo {author} {\bibfnamefont {Y.}~\bibnamefont {Xu}}, \bibinfo {author} {\bibfnamefont {S.}~\bibnamefont {Wang}}, \bibinfo {author} {\bibfnamefont {F.}~\bibnamefont {Qin}}, \bibinfo {author} {\bibfnamefont {Y.}~\bibnamefont {Cao}}, \bibinfo {author} {\bibfnamefont {B.-O.}\ \bibnamefont {Guan}}, \bibinfo {author} {\bibfnamefont {A.}~\bibnamefont {Miroshnichenko}}, \bibinfo {author} {\bibfnamefont {M.}~\bibnamefont {Gu}},\ and\ \bibinfo {author} {\bibfnamefont {X.}~\bibnamefont {Li}},\ }\href {https://doi.org/https://doi.org/10.1016/j.nanoen.2019.05.065} {\bibfield  {journal} {\bibinfo  {journal} {Nano Energy}\ }\textbf {\bibinfo {volume} {62}},\ \bibinfo {pages} {682} (\bibinfo {year} {2019})}\BibitemShut {NoStop}%
\bibitem [{\citenamefont {Kim}\ \emph {et~al.}(2023)\citenamefont {Kim}, \citenamefont {Cho}, \citenamefont {Hong}, \citenamefont {Lee}, \citenamefont {Hwang}, \citenamefont {Kim},\ and\ \citenamefont {Chung}}]{Kim2023ColoursPVDiffractive}%
  \BibitemOpen
  \bibfield  {author} {\bibinfo {author} {\bibfnamefont {W.-J.}\ \bibnamefont {Kim}}, \bibinfo {author} {\bibfnamefont {D.-H.}\ \bibnamefont {Cho}}, \bibinfo {author} {\bibfnamefont {S.-H.}\ \bibnamefont {Hong}}, \bibinfo {author} {\bibfnamefont {W.-J.}\ \bibnamefont {Lee}}, \bibinfo {author} {\bibfnamefont {T.-H.}\ \bibnamefont {Hwang}}, \bibinfo {author} {\bibfnamefont {J.~Y.}\ \bibnamefont {Kim}},\ and\ \bibinfo {author} {\bibfnamefont {Y.-D.}\ \bibnamefont {Chung}},\ }\href {https://doi.org/https://doi.org/10.1016/j.solmat.2023.112392} {\bibfield  {journal} {\bibinfo  {journal} {Solar Energy Materials and Solar Cells}\ }\textbf {\bibinfo {volume} {257}},\ \bibinfo {pages} {112392} (\bibinfo {year} {2023})}\BibitemShut {NoStop}%
\bibitem [{\citenamefont {Lapidas}\ \emph {et~al.}(2024)\citenamefont {Lapidas}, \citenamefont {Cherepakhin}, \citenamefont {Storozhenko}, \citenamefont {Gurevich}, \citenamefont {Zhizhchenko},\ and\ \citenamefont {Kuchmizhak}}]{Lapidas2024PlasmonicEncryption}%
  \BibitemOpen
  \bibfield  {author} {\bibinfo {author} {\bibfnamefont {V.}~\bibnamefont {Lapidas}}, \bibinfo {author} {\bibfnamefont {A.}~\bibnamefont {Cherepakhin}}, \bibinfo {author} {\bibfnamefont {D.}~\bibnamefont {Storozhenko}}, \bibinfo {author} {\bibfnamefont {E.~L.}\ \bibnamefont {Gurevich}}, \bibinfo {author} {\bibfnamefont {A.}~\bibnamefont {Zhizhchenko}},\ and\ \bibinfo {author} {\bibfnamefont {A.~A.}\ \bibnamefont {Kuchmizhak}},\ }\href {https://doi.org/10.1021/acs.nanolett.4c03576} {\bibfield  {journal} {\bibinfo  {journal} {Nano Letters}\ }\textbf {\bibinfo {volume} {24}},\ \bibinfo {pages} {12590} (\bibinfo {year} {2024})}\BibitemShut {NoStop}%
\bibitem [{\citenamefont {Syubaev}\ \emph {et~al.}(2022)\citenamefont {Syubaev}, \citenamefont {Gordeev}, \citenamefont {Modin}, \citenamefont {Terentyev}, \citenamefont {Storozhenko}, \citenamefont {Starikov},\ and\ \citenamefont {Kuchmizhak}}]{Syubaev2022SecurityLabelInfoEncryption}%
  \BibitemOpen
  \bibfield  {author} {\bibinfo {author} {\bibfnamefont {S.}~\bibnamefont {Syubaev}}, \bibinfo {author} {\bibfnamefont {I.}~\bibnamefont {Gordeev}}, \bibinfo {author} {\bibfnamefont {E.}~\bibnamefont {Modin}}, \bibinfo {author} {\bibfnamefont {V.}~\bibnamefont {Terentyev}}, \bibinfo {author} {\bibfnamefont {D.}~\bibnamefont {Storozhenko}}, \bibinfo {author} {\bibfnamefont {S.}~\bibnamefont {Starikov}},\ and\ \bibinfo {author} {\bibfnamefont {A.~A.}\ \bibnamefont {Kuchmizhak}},\ }\href {https://doi.org/10.1039/D2NR04179K} {\bibfield  {journal} {\bibinfo  {journal} {Nanoscale}\ }\textbf {\bibinfo {volume} {14}},\ \bibinfo {pages} {16618} (\bibinfo {year} {2022})}\BibitemShut {NoStop}%
\bibitem [{\citenamefont {Zhou}\ \emph {et~al.}(2024)\citenamefont {Zhou}, \citenamefont {Zhu}, \citenamefont {Cao}, \citenamefont {Wang}, \citenamefont {Kong},\ and\ \citenamefont {Cao}}]{Zhou2024PolarisationSensitiveEncryption}%
  \BibitemOpen
  \bibfield  {author} {\bibinfo {author} {\bibfnamefont {X.}~\bibnamefont {Zhou}}, \bibinfo {author} {\bibfnamefont {H.}~\bibnamefont {Zhu}}, \bibinfo {author} {\bibfnamefont {K.}~\bibnamefont {Cao}}, \bibinfo {author} {\bibfnamefont {Y.}~\bibnamefont {Wang}}, \bibinfo {author} {\bibfnamefont {Y.}~\bibnamefont {Kong}},\ and\ \bibinfo {author} {\bibfnamefont {J.}~\bibnamefont {Cao}},\ }\href {https://doi.org/10.1021/acsami.4c07401} {\bibfield  {journal} {\bibinfo  {journal} {ACS Applied Materials \& Interfaces}\ }\textbf {\bibinfo {volume} {16}},\ \bibinfo {pages} {38404} (\bibinfo {year} {2024})}\BibitemShut {NoStop}%
\bibitem [{\citenamefont {Wang}\ \emph {et~al.}(2021)\citenamefont {Wang}, \citenamefont {Ruan}, \citenamefont {Wang}, \citenamefont {Rezaei}, \citenamefont {Lim}, \citenamefont {Liu}, \citenamefont {Zhang}, \citenamefont {Trisno}, \citenamefont {Chan},\ and\ \citenamefont {Yang}}]{Wang2021Nanopillars}%
  \BibitemOpen
  \bibfield  {author} {\bibinfo {author} {\bibfnamefont {H.}~\bibnamefont {Wang}}, \bibinfo {author} {\bibfnamefont {Q.}~\bibnamefont {Ruan}}, \bibinfo {author} {\bibfnamefont {H.}~\bibnamefont {Wang}}, \bibinfo {author} {\bibfnamefont {S.~D.}\ \bibnamefont {Rezaei}}, \bibinfo {author} {\bibfnamefont {K.~T.~P.}\ \bibnamefont {Lim}}, \bibinfo {author} {\bibfnamefont {H.}~\bibnamefont {Liu}}, \bibinfo {author} {\bibfnamefont {W.}~\bibnamefont {Zhang}}, \bibinfo {author} {\bibfnamefont {J.}~\bibnamefont {Trisno}}, \bibinfo {author} {\bibfnamefont {J.~Y.~E.}\ \bibnamefont {Chan}},\ and\ \bibinfo {author} {\bibfnamefont {J.~K.~W.}\ \bibnamefont {Yang}},\ }\href {https://doi.org/10.1021/acs.nanolett.1c00979} {\bibfield  {journal} {\bibinfo  {journal} {Nano Letters}\ }\textbf {\bibinfo {volume} {21}},\ \bibinfo {pages} {4721} (\bibinfo {year} {2021})}\BibitemShut {NoStop}%
\bibitem [{\citenamefont {Liu}\ \emph {et~al.}(2019)\citenamefont {Liu}, \citenamefont {Zhang}, \citenamefont {Gao}, \citenamefont {Wu}, \citenamefont {Cui},\ and\ \citenamefont {Xu}}]{Liu2019ChiralColour}%
  \BibitemOpen
  \bibfield  {author} {\bibinfo {author} {\bibfnamefont {H.~L.}\ \bibnamefont {Liu}}, \bibinfo {author} {\bibfnamefont {B.}~\bibnamefont {Zhang}}, \bibinfo {author} {\bibfnamefont {T.}~\bibnamefont {Gao}}, \bibinfo {author} {\bibfnamefont {X.}~\bibnamefont {Wu}}, \bibinfo {author} {\bibfnamefont {F.}~\bibnamefont {Cui}},\ and\ \bibinfo {author} {\bibfnamefont {W.}~\bibnamefont {Xu}},\ }\href {https://doi.org/10.1039/C8NR09975H} {\bibfield  {journal} {\bibinfo  {journal} {Nanoscale}\ }\textbf {\bibinfo {volume} {11}},\ \bibinfo {pages} {5506} (\bibinfo {year} {2019})}\BibitemShut {NoStop}%
\bibitem [{\citenamefont {Ng}\ \emph {et~al.}(2019)\citenamefont {Ng}, \citenamefont {Krishnan}, \citenamefont {Dong}, \citenamefont {Ho}, \citenamefont {Liu}, \citenamefont {Ruan}, \citenamefont {Pey},\ and\ \citenamefont {Yang}}]{HongNg2019Microtags}%
  \BibitemOpen
  \bibfield  {author} {\bibinfo {author} {\bibfnamefont {R.~J.~H.}\ \bibnamefont {Ng}}, \bibinfo {author} {\bibfnamefont {R.~V.}\ \bibnamefont {Krishnan}}, \bibinfo {author} {\bibfnamefont {Z.}~\bibnamefont {Dong}}, \bibinfo {author} {\bibfnamefont {J.}~\bibnamefont {Ho}}, \bibinfo {author} {\bibfnamefont {H.}~\bibnamefont {Liu}}, \bibinfo {author} {\bibfnamefont {Q.}~\bibnamefont {Ruan}}, \bibinfo {author} {\bibfnamefont {K.~L.}\ \bibnamefont {Pey}},\ and\ \bibinfo {author} {\bibfnamefont {J.~K.~W.}\ \bibnamefont {Yang}},\ }\href {https://doi.org/10.1364/OME.9.000788} {\bibfield  {journal} {\bibinfo  {journal} {Opt. Mater. Express}\ }\textbf {\bibinfo {volume} {9}},\ \bibinfo {pages} {788} (\bibinfo {year} {2019})}\BibitemShut {NoStop}%
\bibitem [{\citenamefont {Hu}\ \emph {et~al.}(2018)\citenamefont {Hu}, \citenamefont {Lu}, \citenamefont {Cao}, \citenamefont {Zhang}, \citenamefont {Xu}, \citenamefont {Li}, \citenamefont {Gao}, \citenamefont {Cai}, \citenamefont {Guan}, \citenamefont {Qiu},\ and\ \citenamefont {Li}}]{Hu2018LaserSteganography}%
  \BibitemOpen
  \bibfield  {author} {\bibinfo {author} {\bibfnamefont {D.}~\bibnamefont {Hu}}, \bibinfo {author} {\bibfnamefont {Y.}~\bibnamefont {Lu}}, \bibinfo {author} {\bibfnamefont {Y.}~\bibnamefont {Cao}}, \bibinfo {author} {\bibfnamefont {Y.}~\bibnamefont {Zhang}}, \bibinfo {author} {\bibfnamefont {Y.}~\bibnamefont {Xu}}, \bibinfo {author} {\bibfnamefont {W.}~\bibnamefont {Li}}, \bibinfo {author} {\bibfnamefont {F.}~\bibnamefont {Gao}}, \bibinfo {author} {\bibfnamefont {B.}~\bibnamefont {Cai}}, \bibinfo {author} {\bibfnamefont {B.-O.}\ \bibnamefont {Guan}}, \bibinfo {author} {\bibfnamefont {C.-W.}\ \bibnamefont {Qiu}},\ and\ \bibinfo {author} {\bibfnamefont {X.}~\bibnamefont {Li}},\ }\href {https://doi.org/10.1021/acsnano.8b03964} {\bibfield  {journal} {\bibinfo  {journal} {ACS Nano}\ }\textbf {\bibinfo {volume} {12}},\ \bibinfo {pages} {9233} (\bibinfo {year} {2018})}\BibitemShut {NoStop}%
\bibitem [{\citenamefont {Kristensen}\ \emph {et~al.}(2016)\citenamefont {Kristensen}, \citenamefont {Yang}, \citenamefont {Bozhevolnyi}, \citenamefont {Link}, \citenamefont {Nordlander}, \citenamefont {Halas},\ and\ \citenamefont {Mortensen}}]{Kristensen2016PlasmonColourGen}%
  \BibitemOpen
  \bibfield  {author} {\bibinfo {author} {\bibfnamefont {A.}~\bibnamefont {Kristensen}}, \bibinfo {author} {\bibfnamefont {J.~K.~W.}\ \bibnamefont {Yang}}, \bibinfo {author} {\bibfnamefont {S.~I.}\ \bibnamefont {Bozhevolnyi}}, \bibinfo {author} {\bibfnamefont {S.}~\bibnamefont {Link}}, \bibinfo {author} {\bibfnamefont {P.}~\bibnamefont {Nordlander}}, \bibinfo {author} {\bibfnamefont {N.~J.}\ \bibnamefont {Halas}},\ and\ \bibinfo {author} {\bibfnamefont {N.~A.}\ \bibnamefont {Mortensen}},\ }\href {https://doi.org/10.1038/natrevmats.2016.88} {\bibfield  {journal} {\bibinfo  {journal} {Nature Reviews Materials}\ }\textbf {\bibinfo {volume} {2}},\ \bibinfo {pages} {16088} (\bibinfo {year} {2016})}\BibitemShut {NoStop}%
\bibitem [{\citenamefont {Cheng}\ \emph {et~al.}(2015)\citenamefont {Cheng}, \citenamefont {Gao}, \citenamefont {Luk},\ and\ \citenamefont {Yang}}]{Cheng2015PlasmonColourPerfectAbsorption}%
  \BibitemOpen
  \bibfield  {author} {\bibinfo {author} {\bibfnamefont {F.}~\bibnamefont {Cheng}}, \bibinfo {author} {\bibfnamefont {J.}~\bibnamefont {Gao}}, \bibinfo {author} {\bibfnamefont {T.~S.}\ \bibnamefont {Luk}},\ and\ \bibinfo {author} {\bibfnamefont {X.}~\bibnamefont {Yang}},\ }\href {https://doi.org/10.1038/srep11045} {\bibfield  {journal} {\bibinfo  {journal} {Scientific Reports}\ }\textbf {\bibinfo {volume} {5}},\ \bibinfo {pages} {11045} (\bibinfo {year} {2015})}\BibitemShut {NoStop}%
\bibitem [{\citenamefont {Goh}\ \emph {et~al.}(2014)\citenamefont {Goh}, \citenamefont {Zheng}, \citenamefont {Tan}, \citenamefont {Zhang}, \citenamefont {Kumar}, \citenamefont {Qiu},\ and\ \citenamefont {Yang}}]{Goh20143DStereoscopicPlasmonPrint}%
  \BibitemOpen
  \bibfield  {author} {\bibinfo {author} {\bibfnamefont {X.~M.}\ \bibnamefont {Goh}}, \bibinfo {author} {\bibfnamefont {Y.}~\bibnamefont {Zheng}}, \bibinfo {author} {\bibfnamefont {S.~J.}\ \bibnamefont {Tan}}, \bibinfo {author} {\bibfnamefont {L.}~\bibnamefont {Zhang}}, \bibinfo {author} {\bibfnamefont {K.}~\bibnamefont {Kumar}}, \bibinfo {author} {\bibfnamefont {C.-W.}\ \bibnamefont {Qiu}},\ and\ \bibinfo {author} {\bibfnamefont {J.~K.~W.}\ \bibnamefont {Yang}},\ }\href {https://doi.org/10.1038/ncomms6361} {\bibfield  {journal} {\bibinfo  {journal} {Nature Communications}\ }\textbf {\bibinfo {volume} {5}},\ \bibinfo {pages} {5361} (\bibinfo {year} {2014})}\BibitemShut {NoStop}%
\bibitem [{\citenamefont {Franklin}\ \emph {et~al.}(2020)\citenamefont {Franklin}, \citenamefont {He}, \citenamefont {Ortega}, \citenamefont {Safaei}, \citenamefont {Cencillo-Abad}, \citenamefont {Wu},\ and\ \citenamefont {Chanda}}]{Franklin2020SelfAssemblyPlasmonicColours}%
  \BibitemOpen
  \bibfield  {author} {\bibinfo {author} {\bibfnamefont {D.}~\bibnamefont {Franklin}}, \bibinfo {author} {\bibfnamefont {Z.}~\bibnamefont {He}}, \bibinfo {author} {\bibfnamefont {P.~M.}\ \bibnamefont {Ortega}}, \bibinfo {author} {\bibfnamefont {A.}~\bibnamefont {Safaei}}, \bibinfo {author} {\bibfnamefont {P.}~\bibnamefont {Cencillo-Abad}}, \bibinfo {author} {\bibfnamefont {S.-T.}\ \bibnamefont {Wu}},\ and\ \bibinfo {author} {\bibfnamefont {D.}~\bibnamefont {Chanda}},\ }\href {https://doi.org/10.1073/pnas.2001435117} {\bibfield  {journal} {\bibinfo  {journal} {Proceedings of the National Academy of Sciences}\ }\textbf {\bibinfo {volume} {117}},\ \bibinfo {pages} {13350} (\bibinfo {year} {2020})}\BibitemShut {NoStop}%
\bibitem [{\citenamefont {Gramotnev}\ and\ \citenamefont {Bozhevolnyi}(2010)}]{Gramotnev2010Plasmonics}%
  \BibitemOpen
  \bibfield  {author} {\bibinfo {author} {\bibfnamefont {D.~K.}\ \bibnamefont {Gramotnev}}\ and\ \bibinfo {author} {\bibfnamefont {S.~I.}\ \bibnamefont {Bozhevolnyi}},\ }\href {https://doi.org/10.1038/nphoton.2009.282} {\bibfield  {journal} {\bibinfo  {journal} {Nature Photonics}\ }\textbf {\bibinfo {volume} {4}},\ \bibinfo {pages} {83} (\bibinfo {year} {2010})}\BibitemShut {NoStop}%
\bibitem [{\citenamefont {Gramotnev}\ and\ \citenamefont {Bozhevolnyi}(2014)}]{Gramotnev2014Nanofocusing}%
  \BibitemOpen
  \bibfield  {author} {\bibinfo {author} {\bibfnamefont {D.~K.}\ \bibnamefont {Gramotnev}}\ and\ \bibinfo {author} {\bibfnamefont {S.~I.}\ \bibnamefont {Bozhevolnyi}},\ }\href {https://doi.org/10.1038/nphoton.2013.232} {\bibfield  {journal} {\bibinfo  {journal} {Nature Photonics}\ }\textbf {\bibinfo {volume} {8}},\ \bibinfo {pages} {13} (\bibinfo {year} {2014})}\BibitemShut {NoStop}%
\bibitem [{\citenamefont {Kumar}\ \emph {et~al.}(2012)\citenamefont {Kumar}, \citenamefont {Duan}, \citenamefont {Koh}, \citenamefont {Wei},\ and\ \citenamefont {Yang}}]{Kumar2012PrintingDiffractionLimit}%
  \BibitemOpen
  \bibfield  {author} {\bibinfo {author} {\bibfnamefont {K.}~\bibnamefont {Kumar}}, \bibinfo {author} {\bibfnamefont {R.~S.}\ \bibnamefont {Duan}, \bibfnamefont {Huigao;~Hegde}}, \bibinfo {author} {\bibfnamefont {S.~C.~W.}\ \bibnamefont {Koh}}, \bibinfo {author} {\bibfnamefont {J.~N.}\ \bibnamefont {Wei}},\ and\ \bibinfo {author} {\bibfnamefont {J.~K.~W.}\ \bibnamefont {Yang}},\ }\href {https://doi.org/10.1038/nnano.2012.128} {\bibfield  {journal} {\bibinfo  {journal} {Nature Nanotechnology}\ }\textbf {\bibinfo {volume} {7}},\ \bibinfo {pages} {557} (\bibinfo {year} {2012})}\BibitemShut {NoStop}%
\bibitem [{\citenamefont {Roberts}\ \emph {et~al.}(2014)\citenamefont {Roberts}, \citenamefont {Pors}, \citenamefont {Albrektsen},\ and\ \citenamefont {Bozhevolnyi}}]{Roberts2014SubwavelengthPrint}%
  \BibitemOpen
  \bibfield  {author} {\bibinfo {author} {\bibfnamefont {A.~S.}\ \bibnamefont {Roberts}}, \bibinfo {author} {\bibfnamefont {A.}~\bibnamefont {Pors}}, \bibinfo {author} {\bibfnamefont {O.}~\bibnamefont {Albrektsen}},\ and\ \bibinfo {author} {\bibfnamefont {S.~I.}\ \bibnamefont {Bozhevolnyi}},\ }\href {https://doi.org/10.1021/nl404129n} {\bibfield  {journal} {\bibinfo  {journal} {Nano Letters}\ }\textbf {\bibinfo {volume} {14}},\ \bibinfo {pages} {783} (\bibinfo {year} {2014})}\BibitemShut {NoStop}%
\bibitem [{\citenamefont {Todisco}\ \emph {et~al.}(2020)\citenamefont {Todisco}, \citenamefont {Malureanu}, \citenamefont {Wolff}, \citenamefont {Gon\c{c}alves}, \citenamefont {Roberts}, \citenamefont {Mortensen},\ and\ \citenamefont {Tserkezis}}]{todisco2020EPolaritonsSiDisk}%
  \BibitemOpen
  \bibfield  {author} {\bibinfo {author} {\bibfnamefont {F.}~\bibnamefont {Todisco}}, \bibinfo {author} {\bibfnamefont {R.}~\bibnamefont {Malureanu}}, \bibinfo {author} {\bibfnamefont {C.}~\bibnamefont {Wolff}}, \bibinfo {author} {\bibfnamefont {P.~A.~D.}\ \bibnamefont {Gon\c{c}alves}}, \bibinfo {author} {\bibfnamefont {A.~S.}\ \bibnamefont {Roberts}}, \bibinfo {author} {\bibfnamefont {N.~A.}\ \bibnamefont {Mortensen}},\ and\ \bibinfo {author} {\bibfnamefont {C.}~\bibnamefont {Tserkezis}},\ }\href {https://doi.org/10.1515/nanoph-2019-0444} {\bibfield  {journal} {\bibinfo  {journal} {Nanophotonics}\ }\textbf {\bibinfo {volume} {9}},\ \bibinfo {pages} {803} (\bibinfo {year} {2020})}\BibitemShut {NoStop}%
\bibitem [{\citenamefont {Verre}\ \emph {et~al.}(2019)\citenamefont {Verre}, \citenamefont {Baranov}, \citenamefont {Munkhbat}, \citenamefont {Cuadra}, \citenamefont {Käll},\ and\ \citenamefont {Shegai}}]{Verre2019TMDMieNanodisc}%
  \BibitemOpen
  \bibfield  {author} {\bibinfo {author} {\bibfnamefont {R.}~\bibnamefont {Verre}}, \bibinfo {author} {\bibfnamefont {D.~G.}\ \bibnamefont {Baranov}}, \bibinfo {author} {\bibfnamefont {B.}~\bibnamefont {Munkhbat}}, \bibinfo {author} {\bibfnamefont {J.}~\bibnamefont {Cuadra}}, \bibinfo {author} {\bibfnamefont {M.}~\bibnamefont {Käll}},\ and\ \bibinfo {author} {\bibfnamefont {T.}~\bibnamefont {Shegai}},\ }\href {https://doi.org/10.1038/s41565-019-0442-x} {\bibfield  {journal} {\bibinfo  {journal} {Nature Nanotechnology}\ }\textbf {\bibinfo {volume} {14}},\ \bibinfo {pages} {679} (\bibinfo {year} {2019})}\BibitemShut {NoStop}%
\bibitem [{\citenamefont {Evlyukhin}\ \emph {et~al.}(2012)\citenamefont {Evlyukhin}, \citenamefont {Novikov}, \citenamefont {Zywietz}, \citenamefont {Eriksen}, \citenamefont {Reinhardt}, \citenamefont {Bozhevolnyi},\ and\ \citenamefont {Chichkov}}]{evlyukhin_nl12}%
  \BibitemOpen
  \bibfield  {author} {\bibinfo {author} {\bibfnamefont {A.~B.}\ \bibnamefont {Evlyukhin}}, \bibinfo {author} {\bibfnamefont {S.~M.}\ \bibnamefont {Novikov}}, \bibinfo {author} {\bibfnamefont {U.}~\bibnamefont {Zywietz}}, \bibinfo {author} {\bibfnamefont {R.~L.}\ \bibnamefont {Eriksen}}, \bibinfo {author} {\bibfnamefont {C.}~\bibnamefont {Reinhardt}}, \bibinfo {author} {\bibfnamefont {S.~I.}\ \bibnamefont {Bozhevolnyi}},\ and\ \bibinfo {author} {\bibfnamefont {B.~N.}\ \bibnamefont {Chichkov}},\ }\href {https://doi.org/10.1021/nl301594s} {\bibfield  {journal} {\bibinfo  {journal} {Nano Lett.}\ }\textbf {\bibinfo {volume} {12}},\ \bibinfo {pages} {3749} (\bibinfo {year} {2012})}\BibitemShut {NoStop}%
\bibitem [{\citenamefont {Kivshar}(2022)}]{kivshar_nl22}%
  \BibitemOpen
  \bibfield  {author} {\bibinfo {author} {\bibfnamefont {Y.}~\bibnamefont {Kivshar}},\ }\href {https://doi.org/10.1021/acs.nanolett.2c00548} {\bibfield  {journal} {\bibinfo  {journal} {Nano Lett.}\ }\textbf {\bibinfo {volume} {22}},\ \bibinfo {pages} {3513} (\bibinfo {year} {2022})}\BibitemShut {NoStop}%
\bibitem [{\citenamefont {Kuznetsov}\ \emph {et~al.}(2016)\citenamefont {Kuznetsov}, \citenamefont {Miroshnichenko}, \citenamefont {Brongersma}, \citenamefont {Kivshar},\ and\ \citenamefont {Luk’yanchuk}}]{Arseniy2016DielectricResonance}%
  \BibitemOpen
  \bibfield  {author} {\bibinfo {author} {\bibfnamefont {A.~I.}\ \bibnamefont {Kuznetsov}}, \bibinfo {author} {\bibfnamefont {A.~E.}\ \bibnamefont {Miroshnichenko}}, \bibinfo {author} {\bibfnamefont {M.~L.}\ \bibnamefont {Brongersma}}, \bibinfo {author} {\bibfnamefont {Y.~S.}\ \bibnamefont {Kivshar}},\ and\ \bibinfo {author} {\bibfnamefont {B.}~\bibnamefont {Luk’yanchuk}},\ }\href {https://doi.org/10.1126/science.aag2472} {\bibfield  {journal} {\bibinfo  {journal} {Science}\ }\textbf {\bibinfo {volume} {354}},\ \bibinfo {pages} {aag2472} (\bibinfo {year} {2016})}\BibitemShut {NoStop}%
\bibitem [{\citenamefont {Cao}\ \emph {et~al.}(2010)\citenamefont {Cao}, \citenamefont {Fan}, \citenamefont {Barnard}, \citenamefont {Brown},\ and\ \citenamefont {Brongersma}}]{Cao2010SiliconWireColour}%
  \BibitemOpen
  \bibfield  {author} {\bibinfo {author} {\bibfnamefont {L.}~\bibnamefont {Cao}}, \bibinfo {author} {\bibfnamefont {P.}~\bibnamefont {Fan}}, \bibinfo {author} {\bibfnamefont {E.~S.}\ \bibnamefont {Barnard}}, \bibinfo {author} {\bibfnamefont {A.~M.}\ \bibnamefont {Brown}},\ and\ \bibinfo {author} {\bibfnamefont {M.~L.}\ \bibnamefont {Brongersma}},\ }\href {https://doi.org/10.1021/nl1013794} {\bibfield  {journal} {\bibinfo  {journal} {Nano Letters}\ }\textbf {\bibinfo {volume} {10}},\ \bibinfo {pages} {2649} (\bibinfo {year} {2010})}\BibitemShut {NoStop}%
\bibitem [{\citenamefont {Flauraud}\ \emph {et~al.}(2017)\citenamefont {Flauraud}, \citenamefont {Reyes}, \citenamefont {Paniagua-Domínguez}, \citenamefont {Kuznetsov},\ and\ \citenamefont {Brugger}}]{Faluraud2017SiliconFullColourPrints}%
  \BibitemOpen
  \bibfield  {author} {\bibinfo {author} {\bibfnamefont {V.}~\bibnamefont {Flauraud}}, \bibinfo {author} {\bibfnamefont {M.}~\bibnamefont {Reyes}}, \bibinfo {author} {\bibfnamefont {R.}~\bibnamefont {Paniagua-Domínguez}}, \bibinfo {author} {\bibfnamefont {A.~I.}\ \bibnamefont {Kuznetsov}},\ and\ \bibinfo {author} {\bibfnamefont {J.}~\bibnamefont {Brugger}},\ }\href {https://doi.org/10.1021/acsphotonics.6b01021} {\bibfield  {journal} {\bibinfo  {journal} {ACS Photonics}\ }\textbf {\bibinfo {volume} {4}},\ \bibinfo {pages} {1913} (\bibinfo {year} {2017})}\BibitemShut {NoStop}%
\bibitem [{\citenamefont {Yang}\ \emph {et~al.}(2020{\natexlab{a}})\citenamefont {Yang}, \citenamefont {Babicheva}, \citenamefont {Yu}, \citenamefont {Lu}, \citenamefont {Lin},\ and\ \citenamefont {Chen}}]{Yang2020StructuralColoursSi3N4}%
  \BibitemOpen
  \bibfield  {author} {\bibinfo {author} {\bibfnamefont {J.-H.}\ \bibnamefont {Yang}}, \bibinfo {author} {\bibfnamefont {V.~E.}\ \bibnamefont {Babicheva}}, \bibinfo {author} {\bibfnamefont {M.-W.}\ \bibnamefont {Yu}}, \bibinfo {author} {\bibfnamefont {T.-C.}\ \bibnamefont {Lu}}, \bibinfo {author} {\bibfnamefont {T.-R.}\ \bibnamefont {Lin}},\ and\ \bibinfo {author} {\bibfnamefont {K.-P.}\ \bibnamefont {Chen}},\ }\href {https://doi.org/10.1021/acsnano.0c00185} {\bibfield  {journal} {\bibinfo  {journal} {ACS Nano}\ }\textbf {\bibinfo {volume} {14}},\ \bibinfo {pages} {5678} (\bibinfo {year} {2020}{\natexlab{a}})}\BibitemShut {NoStop}%
\bibitem [{\citenamefont {Yang}\ \emph {et~al.}(2020{\natexlab{b}})\citenamefont {Yang}, \citenamefont {Xiao}, \citenamefont {Song}, \citenamefont {Liu}, \citenamefont {Wu}, \citenamefont {Wang}, \citenamefont {Yu}, \citenamefont {Han},\ and\ \citenamefont {Tsai}}]{Yang2020DielectricHighPerformance}%
  \BibitemOpen
  \bibfield  {author} {\bibinfo {author} {\bibfnamefont {W.}~\bibnamefont {Yang}}, \bibinfo {author} {\bibfnamefont {S.}~\bibnamefont {Xiao}}, \bibinfo {author} {\bibfnamefont {Q.}~\bibnamefont {Song}}, \bibinfo {author} {\bibfnamefont {Y.}~\bibnamefont {Liu}}, \bibinfo {author} {\bibfnamefont {Y.}~\bibnamefont {Wu}}, \bibinfo {author} {\bibfnamefont {S.}~\bibnamefont {Wang}}, \bibinfo {author} {\bibfnamefont {J.}~\bibnamefont {Yu}}, \bibinfo {author} {\bibfnamefont {J.}~\bibnamefont {Han}},\ and\ \bibinfo {author} {\bibfnamefont {D.-P.}\ \bibnamefont {Tsai}},\ }\href {https://doi.org/10.1038/s41467-020-15773-0} {\bibfield  {journal} {\bibinfo  {journal} {Nature Communications}\ }\textbf {\bibinfo {volume} {11}},\ \bibinfo {pages} {1864} (\bibinfo {year} {2020}{\natexlab{b}})}\BibitemShut {NoStop}%
\bibitem [{\citenamefont {Manzeli}\ \emph {et~al.}(2017)\citenamefont {Manzeli}, \citenamefont {Ovchinnikov}, \citenamefont {Pasquier}, \citenamefont {Yazyev},\ and\ \citenamefont {Kis}}]{Manzeli20172DTMD}%
  \BibitemOpen
  \bibfield  {author} {\bibinfo {author} {\bibfnamefont {S.}~\bibnamefont {Manzeli}}, \bibinfo {author} {\bibfnamefont {D.}~\bibnamefont {Ovchinnikov}}, \bibinfo {author} {\bibfnamefont {D.}~\bibnamefont {Pasquier}}, \bibinfo {author} {\bibfnamefont {O.~V.}\ \bibnamefont {Yazyev}},\ and\ \bibinfo {author} {\bibfnamefont {A.}~\bibnamefont {Kis}},\ }\href {https://www.nature.com/articles/natrevmats201733} {\bibfield  {journal} {\bibinfo  {journal} {Nature Reviews Materials}\ }\textbf {\bibinfo {volume} {2}},\ \bibinfo {pages} {17033} (\bibinfo {year} {2017})}\BibitemShut {NoStop}%
\bibitem [{\citenamefont {Zhao}\ \emph {et~al.}(2026)\citenamefont {Zhao}, \citenamefont {Alfieri}, \citenamefont {Xia}, \citenamefont {Ge}, \citenamefont {Ge}, \citenamefont {Sun}, \citenamefont {Xie}, \citenamefont {Wang}, \citenamefont {Jariwala}, \citenamefont {Miao},\ and\ \citenamefont {Hu}}]{Zhao2026ExcitonPolaritonDiodes}%
  \BibitemOpen
  \bibfield  {author} {\bibinfo {author} {\bibfnamefont {Q.}~\bibnamefont {Zhao}}, \bibinfo {author} {\bibfnamefont {A.~D.}\ \bibnamefont {Alfieri}}, \bibinfo {author} {\bibfnamefont {M.}~\bibnamefont {Xia}}, \bibinfo {author} {\bibfnamefont {A.}~\bibnamefont {Ge}}, \bibinfo {author} {\bibfnamefont {H.}~\bibnamefont {Ge}}, \bibinfo {author} {\bibfnamefont {L.}~\bibnamefont {Sun}}, \bibinfo {author} {\bibfnamefont {R.}~\bibnamefont {Xie}}, \bibinfo {author} {\bibfnamefont {F.}~\bibnamefont {Wang}}, \bibinfo {author} {\bibfnamefont {D.}~\bibnamefont {Jariwala}}, \bibinfo {author} {\bibfnamefont {J.}~\bibnamefont {Miao}},\ and\ \bibinfo {author} {\bibfnamefont {W.}~\bibnamefont {Hu}},\ }\href {https://doi.org/10.1038/s41467-026-68312-8} {\bibfield  {journal} {\bibinfo  {journal} {Nature Communications}\ }\textbf {\bibinfo {volume} {17}},\ \bibinfo {pages} {1607} (\bibinfo {year} {2026})}\BibitemShut {NoStop}%
\bibitem [{\citenamefont {Ermolaev}\ \emph {et~al.}(2021)\citenamefont {Ermolaev}, \citenamefont {Grudinin}, \citenamefont {Stebunov}, \citenamefont {Voronin}, \citenamefont {Kravets}, \citenamefont {Duan}, \citenamefont {Mazitov}, \citenamefont {Tselikov}, \citenamefont {Bylinkin}, \citenamefont {Yakubovsky}, \citenamefont {Novikov}, \citenamefont {Baranov}, \citenamefont {Nikitin}, \citenamefont {Kruglov}, \citenamefont {Shegai}, \citenamefont {Alonso-González}, \citenamefont {Grigorenko}, \citenamefont {Arsenin}, \citenamefont {Novoselov},\ and\ \citenamefont {Volkov}}]{Ermolaev2021GiantAnisotropy}%
  \BibitemOpen
  \bibfield  {author} {\bibinfo {author} {\bibfnamefont {G.~A.}\ \bibnamefont {Ermolaev}}, \bibinfo {author} {\bibfnamefont {D.~V.}\ \bibnamefont {Grudinin}}, \bibinfo {author} {\bibfnamefont {Y.~V.}\ \bibnamefont {Stebunov}}, \bibinfo {author} {\bibfnamefont {K.~V.}\ \bibnamefont {Voronin}}, \bibinfo {author} {\bibfnamefont {V.~G.}\ \bibnamefont {Kravets}}, \bibinfo {author} {\bibfnamefont {J.}~\bibnamefont {Duan}}, \bibinfo {author} {\bibfnamefont {A.~B.}\ \bibnamefont {Mazitov}}, \bibinfo {author} {\bibfnamefont {G.~I.}\ \bibnamefont {Tselikov}}, \bibinfo {author} {\bibfnamefont {A.}~\bibnamefont {Bylinkin}}, \bibinfo {author} {\bibfnamefont {D.~I.}\ \bibnamefont {Yakubovsky}}, \bibinfo {author} {\bibfnamefont {S.~M.}\ \bibnamefont {Novikov}}, \bibinfo {author} {\bibfnamefont {D.~G.}\ \bibnamefont {Baranov}}, \bibinfo {author} {\bibfnamefont {A.~Y.}\ \bibnamefont {Nikitin}}, \bibinfo {author} {\bibfnamefont {I.~A.}\ \bibnamefont {Kruglov}}, \bibinfo {author} {\bibfnamefont {T.}~\bibnamefont {Shegai}},
  \bibinfo {author} {\bibfnamefont {P.}~\bibnamefont {Alonso-González}}, \bibinfo {author} {\bibfnamefont {A.~N.}\ \bibnamefont {Grigorenko}}, \bibinfo {author} {\bibfnamefont {A.~V.}\ \bibnamefont {Arsenin}}, \bibinfo {author} {\bibfnamefont {K.~S.}\ \bibnamefont {Novoselov}},\ and\ \bibinfo {author} {\bibfnamefont {V.~S.}\ \bibnamefont {Volkov}},\ }\href@noop {} {\bibfield  {journal} {\bibinfo  {journal} {Nature Communications}\ }\textbf {\bibinfo {volume} {12}},\ \bibinfo {pages} {854} (\bibinfo {year} {2021})}\BibitemShut {NoStop}%
\bibitem [{\citenamefont {Neupane}\ \emph {et~al.}(2019)\citenamefont {Neupane}, \citenamefont {Zhou}, \citenamefont {Chen}, \citenamefont {Yildirim}, \citenamefont {Zhang},\ and\ \citenamefont {Lu}}]{Neupane2019Anisotropy}%
  \BibitemOpen
  \bibfield  {author} {\bibinfo {author} {\bibfnamefont {G.~P.}\ \bibnamefont {Neupane}}, \bibinfo {author} {\bibfnamefont {K.}~\bibnamefont {Zhou}}, \bibinfo {author} {\bibfnamefont {S.}~\bibnamefont {Chen}}, \bibinfo {author} {\bibfnamefont {T.}~\bibnamefont {Yildirim}}, \bibinfo {author} {\bibfnamefont {P.}~\bibnamefont {Zhang}},\ and\ \bibinfo {author} {\bibfnamefont {Y.}~\bibnamefont {Lu}},\ }\href {https://doi.org/https://doi.org/10.1002/smll.201804733} {\bibfield  {journal} {\bibinfo  {journal} {Small}\ }\textbf {\bibinfo {volume} {15}},\ \bibinfo {pages} {1804733} (\bibinfo {year} {2019})}\BibitemShut {NoStop}%
\bibitem [{\citenamefont {Munkhbat}\ \emph {et~al.}(2022)\citenamefont {Munkhbat}, \citenamefont {Wróbel}, \citenamefont {Antosiewicz},\ and\ \citenamefont {Shegai}}]{Munkhbat2022Anisotropy}%
  \BibitemOpen
  \bibfield  {author} {\bibinfo {author} {\bibfnamefont {B.}~\bibnamefont {Munkhbat}}, \bibinfo {author} {\bibfnamefont {P.}~\bibnamefont {Wróbel}}, \bibinfo {author} {\bibfnamefont {T.~J.}\ \bibnamefont {Antosiewicz}},\ and\ \bibinfo {author} {\bibfnamefont {T.~O.}\ \bibnamefont {Shegai}},\ }\href {https://doi.org/10.1021/acsphotonics.2c00433} {\bibfield  {journal} {\bibinfo  {journal} {ACS Photonics}\ }\textbf {\bibinfo {volume} {9}},\ \bibinfo {pages} {2398} (\bibinfo {year} {2022})}\BibitemShut {NoStop}%
\bibitem [{\citenamefont {Wang}\ \emph {et~al.}(2016)\citenamefont {Wang}, \citenamefont {Sun}, \citenamefont {Zhang}, \citenamefont {Chen}, \citenamefont {Shen},\ and\ \citenamefont {Lu}}]{Wang2016ExcitonCouplingWS2}%
  \BibitemOpen
  \bibfield  {author} {\bibinfo {author} {\bibfnamefont {Q.}~\bibnamefont {Wang}}, \bibinfo {author} {\bibfnamefont {L.}~\bibnamefont {Sun}}, \bibinfo {author} {\bibfnamefont {B.}~\bibnamefont {Zhang}}, \bibinfo {author} {\bibfnamefont {C.}~\bibnamefont {Chen}}, \bibinfo {author} {\bibfnamefont {X.}~\bibnamefont {Shen}},\ and\ \bibinfo {author} {\bibfnamefont {W.}~\bibnamefont {Lu}},\ }\href@noop {} {\bibfield  {journal} {\bibinfo  {journal} {Optics Express}\ }\textbf {\bibinfo {volume} {24}},\ \bibinfo {pages} {7151} (\bibinfo {year} {2016})}\BibitemShut {NoStop}%
\bibitem [{\citenamefont {Munkhbat}\ \emph {et~al.}(2019)\citenamefont {Munkhbat}, \citenamefont {Baranov}, \citenamefont {St{\"u}hrenberg}, \citenamefont {Wersäll}, \citenamefont {Bisht},\ and\ \citenamefont {Shegai}}]{Munkhbat2019PlexcitonsTMD}%
  \BibitemOpen
  \bibfield  {author} {\bibinfo {author} {\bibfnamefont {B.}~\bibnamefont {Munkhbat}}, \bibinfo {author} {\bibfnamefont {D.~G.}\ \bibnamefont {Baranov}}, \bibinfo {author} {\bibfnamefont {M.}~\bibnamefont {St{\"u}hrenberg}}, \bibinfo {author} {\bibfnamefont {M.}~\bibnamefont {Wersäll}}, \bibinfo {author} {\bibfnamefont {A.}~\bibnamefont {Bisht}},\ and\ \bibinfo {author} {\bibfnamefont {T.}~\bibnamefont {Shegai}},\ }\href {https://doi.org/10.1021/acsphotonics.8b01194} {\bibfield  {journal} {\bibinfo  {journal} {ACS Photonics}\ }\textbf {\bibinfo {volume} {6}},\ \bibinfo {pages} {139} (\bibinfo {year} {2019})}\BibitemShut {NoStop}%
\bibitem [{\citenamefont {Munkhbat}\ \emph {et~al.}(2023)\citenamefont {Munkhbat}, \citenamefont {Küçüköz}, \citenamefont {Baranov}, \citenamefont {Antosiewicz},\ and\ \citenamefont {Shegai}}]{Munkhbat2023NTMD}%
  \BibitemOpen
  \bibfield  {author} {\bibinfo {author} {\bibfnamefont {B.}~\bibnamefont {Munkhbat}}, \bibinfo {author} {\bibfnamefont {B.}~\bibnamefont {Küçüköz}}, \bibinfo {author} {\bibfnamefont {D.~G.}\ \bibnamefont {Baranov}}, \bibinfo {author} {\bibfnamefont {T.~J.}\ \bibnamefont {Antosiewicz}},\ and\ \bibinfo {author} {\bibfnamefont {T.~O.}\ \bibnamefont {Shegai}},\ }\href {https://onlinelibrary.wiley.com/doi/10.1002/lpor.202200057} {\bibfield  {journal} {\bibinfo  {journal} {Laser \& {P}hotonics {R}eviews}\ }\textbf {\bibinfo {volume} {17}},\ \bibinfo {pages} {2200057} (\bibinfo {year} {2023})}\BibitemShut {NoStop}%
\bibitem [{\citenamefont {Datta}\ \emph {et~al.}(2020)\citenamefont {Datta}, \citenamefont {Chae}, \citenamefont {Bhatt}, \citenamefont {Tadayon}, \citenamefont {Li}, \citenamefont {Yu}, \citenamefont {Park}, \citenamefont {Park}, \citenamefont {Cao}, \citenamefont {Basov}, \citenamefont {Hone},\ and\ \citenamefont {Lipson}}]{Datta2020LowlossPlatform}%
  \BibitemOpen
  \bibfield  {author} {\bibinfo {author} {\bibfnamefont {I.}~\bibnamefont {Datta}}, \bibinfo {author} {\bibfnamefont {S.~H.}\ \bibnamefont {Chae}}, \bibinfo {author} {\bibfnamefont {G.~R.}\ \bibnamefont {Bhatt}}, \bibinfo {author} {\bibfnamefont {M.~A.}\ \bibnamefont {Tadayon}}, \bibinfo {author} {\bibfnamefont {B.}~\bibnamefont {Li}}, \bibinfo {author} {\bibfnamefont {Y.}~\bibnamefont {Yu}}, \bibinfo {author} {\bibfnamefont {C.}~\bibnamefont {Park}}, \bibinfo {author} {\bibfnamefont {J.}~\bibnamefont {Park}}, \bibinfo {author} {\bibfnamefont {L.}~\bibnamefont {Cao}}, \bibinfo {author} {\bibfnamefont {D.~N.}\ \bibnamefont {Basov}}, \bibinfo {author} {\bibfnamefont {J.}~\bibnamefont {Hone}},\ and\ \bibinfo {author} {\bibfnamefont {M.}~\bibnamefont {Lipson}},\ }\href {https://doi.org/10.1038/s41566-020-0590-4} {\bibfield  {journal} {\bibinfo  {journal} {Nature Photonics}\ }\textbf {\bibinfo {volume} {14}},\ \bibinfo {pages} {256} (\bibinfo {year} {2020})}\BibitemShut {NoStop}%
\bibitem [{\citenamefont {Born}\ \emph {et~al.}(1999)\citenamefont {Born}, \citenamefont {Wolf}, \citenamefont {Bhatia}, \citenamefont {Clemmow}, \citenamefont {Gabor}, \citenamefont {Stokes}, \citenamefont {Taylor}, \citenamefont {Wayman},\ and\ \citenamefont {Wilcock}}]{BornWolfPrinciplesOptics}%
  \BibitemOpen
  \bibfield  {author} {\bibinfo {author} {\bibfnamefont {M.}~\bibnamefont {Born}}, \bibinfo {author} {\bibfnamefont {E.}~\bibnamefont {Wolf}}, \bibinfo {author} {\bibfnamefont {A.~B.}\ \bibnamefont {Bhatia}}, \bibinfo {author} {\bibfnamefont {P.~C.}\ \bibnamefont {Clemmow}}, \bibinfo {author} {\bibfnamefont {D.}~\bibnamefont {Gabor}}, \bibinfo {author} {\bibfnamefont {A.~R.}\ \bibnamefont {Stokes}}, \bibinfo {author} {\bibfnamefont {A.~M.}\ \bibnamefont {Taylor}}, \bibinfo {author} {\bibfnamefont {P.~A.}\ \bibnamefont {Wayman}},\ and\ \bibinfo {author} {\bibfnamefont {W.~L.}\ \bibnamefont {Wilcock}},\ }\href {https://www.cambridge.org/core/books/principles-of-optics/D12868B8AE26B83D6D3C2193E94FFC32} {\emph {\bibinfo {title} {Principles of Optics: Electromagnetic Theory of Propagation, Interference and Diffraction of Light}}}\ (\bibinfo  {publisher} {Cambridge University Press},\ \bibinfo {year} {1999})\BibitemShut {NoStop}%
\bibitem [{\citenamefont {Novotny}\ and\ \citenamefont {Hecht}(2012)}]{NovotnyPrinciplesNanoOptics}%
  \BibitemOpen
  \bibfield  {author} {\bibinfo {author} {\bibfnamefont {L.}~\bibnamefont {Novotny}}\ and\ \bibinfo {author} {\bibfnamefont {B.}~\bibnamefont {Hecht}},\ }\href@noop {} {\emph {\bibinfo {title} {Principles of Nano-Optics}}},\ \bibinfo {edition} {2nd}\ ed.\ (\bibinfo  {publisher} {Cambridge University Press},\ \bibinfo {year} {2012})\BibitemShut {NoStop}%
\bibitem [{\citenamefont {Waterman}(1971)}]{Waterman1971ScatteringMatrix}%
  \BibitemOpen
  \bibfield  {author} {\bibinfo {author} {\bibfnamefont {P.~C.}\ \bibnamefont {Waterman}},\ }\href {https://doi.org/10.1103/PhysRevD.3.825} {\bibfield  {journal} {\bibinfo  {journal} {Phys. Rev. D}\ }\textbf {\bibinfo {volume} {3}},\ \bibinfo {pages} {825} (\bibinfo {year} {1971})}\BibitemShut {NoStop}%
\bibitem [{\citenamefont {Mishchenko}\ \emph {et~al.}(2002)\citenamefont {Mishchenko}, \citenamefont {Travis},\ and\ \citenamefont {Lacis}}]{MishchenkoLightBySmallParticles}%
  \BibitemOpen
  \bibfield  {author} {\bibinfo {author} {\bibfnamefont {M.}~\bibnamefont {Mishchenko}}, \bibinfo {author} {\bibfnamefont {L.}~\bibnamefont {Travis}},\ and\ \bibinfo {author} {\bibfnamefont {A.}~\bibnamefont {Lacis}},\ }\href@noop {} {\emph {\bibinfo {title} {Scattering, Absorption, and Emission of Light by Small Particles}}}\ (\bibinfo  {publisher} {Cambridge University Press},\ \bibinfo {year} {2002})\BibitemShut {NoStop}%
\bibitem [{\citenamefont {Mie}(1908)}]{Mie1908}%
  \BibitemOpen
  \bibfield  {author} {\bibinfo {author} {\bibfnamefont {G.}~\bibnamefont {Mie}},\ }\href@noop {} {\bibfield  {journal} {\bibinfo  {journal} {Ann. Phys.}\ }\textbf {\bibinfo {volume} {330}},\ \bibinfo {pages} {377} (\bibinfo {year} {1908})}\BibitemShut {NoStop}%
\bibitem [{\citenamefont {Alvarez-Serrano}\ \emph {et~al.}(2024)\citenamefont {Alvarez-Serrano}, \citenamefont {Deop-Ruano}, \citenamefont {Aglieri}, \citenamefont {Toma},\ and\ \citenamefont {Manjavacas}}]{Alvarez-Serrano2024BipartiteArrays}%
  \BibitemOpen
  \bibfield  {author} {\bibinfo {author} {\bibfnamefont {J.~J.}\ \bibnamefont {Alvarez-Serrano}}, \bibinfo {author} {\bibfnamefont {J.~R.}\ \bibnamefont {Deop-Ruano}}, \bibinfo {author} {\bibfnamefont {V.}~\bibnamefont {Aglieri}}, \bibinfo {author} {\bibfnamefont {A.}~\bibnamefont {Toma}},\ and\ \bibinfo {author} {\bibfnamefont {A.}~\bibnamefont {Manjavacas}},\ }\href {https://doi.org/10.1021/acsphotonics.3c01535} {\bibfield  {journal} {\bibinfo  {journal} {ACS {P}hotonics}\ }\textbf {\bibinfo {volume} {11}},\ \bibinfo {pages} {301} (\bibinfo {year} {2024})}\BibitemShut {NoStop}%
\bibitem [{\citenamefont {Manjavacas}\ \emph {et~al.}(2019)\citenamefont {Manjavacas}, \citenamefont {Zundel},\ and\ \citenamefont {Sanders}}]{Manjavacas2019NearFieldNanoArray}%
  \BibitemOpen
  \bibfield  {author} {\bibinfo {author} {\bibfnamefont {A.}~\bibnamefont {Manjavacas}}, \bibinfo {author} {\bibfnamefont {L.}~\bibnamefont {Zundel}},\ and\ \bibinfo {author} {\bibfnamefont {S.}~\bibnamefont {Sanders}},\ }\href {https://doi.org/10.1021/acsnano.9b05031} {\bibfield  {journal} {\bibinfo  {journal} {ACS {N}ano}\ }\textbf {\bibinfo {volume} {13}},\ \bibinfo {pages} {10682} (\bibinfo {year} {2019})}\BibitemShut {NoStop}%
\bibitem [{\citenamefont {Beutel}\ \emph {et~al.}(2024)\citenamefont {Beutel}, \citenamefont {Fernandez-Corbaton},\ and\ \citenamefont {Rockstuhl}}]{Beutel2024Treams}%
  \BibitemOpen
  \bibfield  {author} {\bibinfo {author} {\bibfnamefont {D.}~\bibnamefont {Beutel}}, \bibinfo {author} {\bibfnamefont {I.}~\bibnamefont {Fernandez-Corbaton}},\ and\ \bibinfo {author} {\bibfnamefont {C.}~\bibnamefont {Rockstuhl}},\ }\href {https://www.sciencedirect.com/science/article/pii/S0010465523004216} {\bibfield  {journal} {\bibinfo  {journal} {Computer Physics Communications}\ }\textbf {\bibinfo {volume} {297}},\ \bibinfo {pages} {109076} (\bibinfo {year} {2024})}\BibitemShut {NoStop}%
\bibitem [{\citenamefont {Malacara}\ and\ \citenamefont {{Society of Photo-optical Instrumentation Engineers}}(2011)}]{MalacaraColorimetry}%
  \BibitemOpen
  \bibfield  {author} {\bibinfo {author} {\bibfnamefont {D.}~\bibnamefont {Malacara}}\ and\ \bibinfo {author} {\bibnamefont {{Society of Photo-optical Instrumentation Engineers}}},\ }\href {https://books.google.dk/books?id=xDU4YgEACAAJ} {\emph {\bibinfo {title} {Color {V}ision and {C}olorimetry: {T}heory and {A}pplications}}},\ SPIE Press monograph\ (\bibinfo  {publisher} {SPIE},\ \bibinfo {year} {2011})\BibitemShut {NoStop}%
\bibitem [{\citenamefont {CIE}(2019{\natexlab{a}})}]{CIE2019ColourMatch1931}%
  \BibitemOpen
  \bibfield  {author} {\bibinfo {author} {\bibnamefont {CIE}},\ }\bibfield  {journal} {\bibinfo  {journal} {International Commission on Illumination (CIE)}\ }\href {https://doi.org/10.25039/CIE.DS.xvudnb9b} {10.25039/CIE.DS.xvudnb9b} (\bibinfo {year} {2019}{\natexlab{a}})\BibitemShut {NoStop}%
\bibitem [{\citenamefont {CIE}(2019{\natexlab{b}})}]{CIE2019IlluminantD65}%
  \BibitemOpen
  \bibfield  {author} {\bibinfo {author} {\bibnamefont {CIE}},\ }\bibfield  {journal} {\bibinfo  {journal} {International Commission on Illumination (CIE)}\ }\href {https://doi.org/10.25039/CIE.DS.hjfjmt59} {10.25039/CIE.DS.hjfjmt59} (\bibinfo {year} {2019}{\natexlab{b}})\BibitemShut {NoStop}%
\bibitem [{\citenamefont {Sharma}\ \emph {et~al.}(2005)\citenamefont {Sharma}, \citenamefont {Wu},\ and\ \citenamefont {Dalal}}]{Sharma2005CIED2000ColorDiff}%
  \BibitemOpen
  \bibfield  {author} {\bibinfo {author} {\bibfnamefont {G.}~\bibnamefont {Sharma}}, \bibinfo {author} {\bibfnamefont {W.}~\bibnamefont {Wu}},\ and\ \bibinfo {author} {\bibfnamefont {E.~N.}\ \bibnamefont {Dalal}},\ }\href {https://doi.org/https://doi.org/10.1002/col.20070} {\bibfield  {journal} {\bibinfo  {journal} {Color Research \& Application}\ }\textbf {\bibinfo {volume} {30}},\ \bibinfo {pages} {21} (\bibinfo {year} {2005})}\BibitemShut {NoStop}%
\bibitem [{\citenamefont {Luo}\ \emph {et~al.}(2001)\citenamefont {Luo}, \citenamefont {Cui},\ and\ \citenamefont {Rigg}}]{Luo2001DevelopmentCIED2000}%
  \BibitemOpen
  \bibfield  {author} {\bibinfo {author} {\bibfnamefont {M.~R.}\ \bibnamefont {Luo}}, \bibinfo {author} {\bibfnamefont {G.}~\bibnamefont {Cui}},\ and\ \bibinfo {author} {\bibfnamefont {B.}~\bibnamefont {Rigg}},\ }\href {https://doi.org/https://doi.org/10.1002/col.1049} {\bibfield  {journal} {\bibinfo  {journal} {Color Research \& Application}\ }\textbf {\bibinfo {volume} {26}},\ \bibinfo {pages} {340} (\bibinfo {year} {2001})}\BibitemShut {NoStop}%
\bibitem [{\citenamefont {Mokrzycki}\ and\ \citenamefont {Tatol}(2011)}]{Mokrzycki2011DeltaE}%
  \BibitemOpen
  \bibfield  {author} {\bibinfo {author} {\bibfnamefont {W.}~\bibnamefont {Mokrzycki}}\ and\ \bibinfo {author} {\bibfnamefont {M.}~\bibnamefont {Tatol}},\ }\href@noop {} {\bibfield  {journal} {\bibinfo  {journal} {Machine Graphics and Vision}\ }\textbf {\bibinfo {volume} {20}},\ \bibinfo {pages} {383} (\bibinfo {year} {2011})}\BibitemShut {NoStop}%
\bibitem [{\citenamefont {Hohenester}(2019)}]{HohenesterNanoQuantumOptics}%
  \BibitemOpen
  \bibfield  {author} {\bibinfo {author} {\bibfnamefont {U.}~\bibnamefont {Hohenester}},\ }\href {https://link.springer.com/book/10.1007/978-3-030-30504-8} {\emph {\bibinfo {title} {{N}ano and {Q}uantum {O}ptics: {A}n {I}ntroduction to {B}asic {P}rinciples and {T}heory}}},\ {G}raduate {T}exts in {P}hysics\ (\bibinfo  {publisher} {{S}pringer {I}nternational {P}ublishing},\ \bibinfo {year} {2019})\BibitemShut {NoStop}%
\bibitem [{\citenamefont {Vyshnevyy}\ \emph {et~al.}(2023)\citenamefont {Vyshnevyy}, \citenamefont {Ermolaev}, \citenamefont {Grudinin}, \citenamefont {Voronin}, \citenamefont {Kharichkin}, \citenamefont {Mazitov}, \citenamefont {Kruglov}, \citenamefont {Yakubovsky}, \citenamefont {Mishra}, \citenamefont {Kirtaev}, \citenamefont {Arsenin}, \citenamefont {Novoselov}, \citenamefont {Martin-Moreno},\ and\ \citenamefont {Volkov}}]{Vyshnevyy2023WS2}%
  \BibitemOpen
  \bibfield  {author} {\bibinfo {author} {\bibfnamefont {A.~A.}\ \bibnamefont {Vyshnevyy}}, \bibinfo {author} {\bibfnamefont {G.~A.}\ \bibnamefont {Ermolaev}}, \bibinfo {author} {\bibfnamefont {D.~V.}\ \bibnamefont {Grudinin}}, \bibinfo {author} {\bibfnamefont {K.~V.}\ \bibnamefont {Voronin}}, \bibinfo {author} {\bibfnamefont {I.}~\bibnamefont {Kharichkin}}, \bibinfo {author} {\bibfnamefont {A.}~\bibnamefont {Mazitov}}, \bibinfo {author} {\bibfnamefont {I.~A.}\ \bibnamefont {Kruglov}}, \bibinfo {author} {\bibfnamefont {D.~I.}\ \bibnamefont {Yakubovsky}}, \bibinfo {author} {\bibfnamefont {P.}~\bibnamefont {Mishra}}, \bibinfo {author} {\bibfnamefont {R.~V.}\ \bibnamefont {Kirtaev}}, \bibinfo {author} {\bibfnamefont {A.~V.}\ \bibnamefont {Arsenin}}, \bibinfo {author} {\bibfnamefont {K.~S.}\ \bibnamefont {Novoselov}}, \bibinfo {author} {\bibfnamefont {L.}~\bibnamefont {Martin-Moreno}},\ and\ \bibinfo {author} {\bibfnamefont {V.~S.}\ \bibnamefont {Volkov}},\ }\href {https://doi.org/10.1021/acs.nanolett.3c02051}
  {\bibfield  {journal} {\bibinfo  {journal} {Nano Letters}\ }\textbf {\bibinfo {volume} {23}},\ \bibinfo {pages} {8057} (\bibinfo {year} {2023})}\BibitemShut {NoStop}%
\bibitem [{\citenamefont {Ermolaev}\ \emph {et~al.}(2020)\citenamefont {Ermolaev}, \citenamefont {Stebunov}, \citenamefont {Vyshnevyy}, \citenamefont {Tatarkin}, \citenamefont {Yakubovsky}, \citenamefont {Novikov}, \citenamefont {Baranov}, \citenamefont {Shegai}, \citenamefont {Nikitin}, \citenamefont {Arsenin},\ and\ \citenamefont {Volkov}}]{Ermolaev2020MoS2}%
  \BibitemOpen
  \bibfield  {author} {\bibinfo {author} {\bibfnamefont {G.~A.}\ \bibnamefont {Ermolaev}}, \bibinfo {author} {\bibfnamefont {Y.~V.}\ \bibnamefont {Stebunov}}, \bibinfo {author} {\bibfnamefont {A.~A.}\ \bibnamefont {Vyshnevyy}}, \bibinfo {author} {\bibfnamefont {D.~E.}\ \bibnamefont {Tatarkin}}, \bibinfo {author} {\bibfnamefont {D.~I.}\ \bibnamefont {Yakubovsky}}, \bibinfo {author} {\bibfnamefont {S.~M.}\ \bibnamefont {Novikov}}, \bibinfo {author} {\bibfnamefont {D.~G.}\ \bibnamefont {Baranov}}, \bibinfo {author} {\bibfnamefont {T.}~\bibnamefont {Shegai}}, \bibinfo {author} {\bibfnamefont {A.~Y.}\ \bibnamefont {Nikitin}}, \bibinfo {author} {\bibfnamefont {A.~V.}\ \bibnamefont {Arsenin}},\ and\ \bibinfo {author} {\bibfnamefont {V.~S.}\ \bibnamefont {Volkov}},\ }\href {https://doi.org/10.1038/s41699-020-0155-x} {\bibfield  {journal} {\bibinfo  {journal} {npj 2D Materials and Applications}\ }\textbf {\bibinfo {volume} {4}},\ \bibinfo {pages} {21} (\bibinfo {year} {2020})}\BibitemShut {NoStop}%
\bibitem [{\citenamefont {Zotev}\ \emph {et~al.}(2023)\citenamefont {Zotev}, \citenamefont {Wang}, \citenamefont {Andres-Penares}, \citenamefont {Severs-Millard}, \citenamefont {Randerson}, \citenamefont {Hu}, \citenamefont {Sortino}, \citenamefont {Louca}, \citenamefont {Brotons-Gisbert}, \citenamefont {Huq}, \citenamefont {Vezzoli}, \citenamefont {Sapienza}, \citenamefont {Krauss}, \citenamefont {Gerardot},\ and\ \citenamefont {Tartakovskii}}]{Zotev2023VanDerWaals}%
  \BibitemOpen
  \bibfield  {author} {\bibinfo {author} {\bibfnamefont {P.~G.}\ \bibnamefont {Zotev}}, \bibinfo {author} {\bibfnamefont {Y.}~\bibnamefont {Wang}}, \bibinfo {author} {\bibfnamefont {D.}~\bibnamefont {Andres-Penares}}, \bibinfo {author} {\bibfnamefont {T.}~\bibnamefont {Severs-Millard}}, \bibinfo {author} {\bibfnamefont {S.}~\bibnamefont {Randerson}}, \bibinfo {author} {\bibfnamefont {X.}~\bibnamefont {Hu}}, \bibinfo {author} {\bibfnamefont {L.}~\bibnamefont {Sortino}}, \bibinfo {author} {\bibfnamefont {C.}~\bibnamefont {Louca}}, \bibinfo {author} {\bibfnamefont {M.}~\bibnamefont {Brotons-Gisbert}}, \bibinfo {author} {\bibfnamefont {T.}~\bibnamefont {Huq}}, \bibinfo {author} {\bibfnamefont {S.}~\bibnamefont {Vezzoli}}, \bibinfo {author} {\bibfnamefont {R.}~\bibnamefont {Sapienza}}, \bibinfo {author} {\bibfnamefont {T.~F.}\ \bibnamefont {Krauss}}, \bibinfo {author} {\bibfnamefont {B.~D.}\ \bibnamefont {Gerardot}},\ and\ \bibinfo {author} {\bibfnamefont {A.~I.}\ \bibnamefont {Tartakovskii}},\ }\href
  {https://onlinelibrary.wiley.com/doi/abs/10.1002/lpor.202200957} {\bibfield  {journal} {\bibinfo  {journal} {Laser \& Photonics Reviews}\ }\textbf {\bibinfo {volume} {17}},\ \bibinfo {pages} {2200957} (\bibinfo {year} {2023})}\BibitemShut {NoStop}%
\bibitem [{\citenamefont {Zhao}\ \emph {et~al.}(2017)\citenamefont {Zhao}, \citenamefont {Guo}, \citenamefont {Si}, \citenamefont {Ren}, \citenamefont {Bai},\ and\ \citenamefont {Xu}}]{Zhao2017DielectricProperties}%
  \BibitemOpen
  \bibfield  {author} {\bibinfo {author} {\bibfnamefont {Q.}~\bibnamefont {Zhao}}, \bibinfo {author} {\bibfnamefont {Y.}~\bibnamefont {Guo}}, \bibinfo {author} {\bibfnamefont {K.}~\bibnamefont {Si}}, \bibinfo {author} {\bibfnamefont {Z.}~\bibnamefont {Ren}}, \bibinfo {author} {\bibfnamefont {J.}~\bibnamefont {Bai}},\ and\ \bibinfo {author} {\bibfnamefont {X.}~\bibnamefont {Xu}},\ }\href {https://onlinelibrary.wiley.com/doi/abs/10.1002/pssb.201700033} {\bibfield  {journal} {\bibinfo  {journal} {Physica Status Solidi (b)}\ }\textbf {\bibinfo {volume} {254}},\ \bibinfo {pages} {1700033} (\bibinfo {year} {2017})}\BibitemShut {NoStop}%
\bibitem [{\citenamefont {Li}\ \emph {et~al.}(2024)\citenamefont {Li}, \citenamefont {Wrzesińska-Lashkova}, \citenamefont {Deconinck}, \citenamefont {Göbel}, \citenamefont {Vaynzof}, \citenamefont {Lesnyak},\ and\ \citenamefont {Eychmüller}}]{Li2024ColloidalSynethesisTMD}%
  \BibitemOpen
  \bibfield  {author} {\bibinfo {author} {\bibfnamefont {J.}~\bibnamefont {Li}}, \bibinfo {author} {\bibfnamefont {A.}~\bibnamefont {Wrzesińska-Lashkova}}, \bibinfo {author} {\bibfnamefont {M.}~\bibnamefont {Deconinck}}, \bibinfo {author} {\bibfnamefont {M.}~\bibnamefont {Göbel}}, \bibinfo {author} {\bibfnamefont {Y.}~\bibnamefont {Vaynzof}}, \bibinfo {author} {\bibfnamefont {V.}~\bibnamefont {Lesnyak}},\ and\ \bibinfo {author} {\bibfnamefont {A.}~\bibnamefont {Eychmüller}},\ }\href {https://doi.org/10.1021/acsami.4c04968} {\bibfield  {journal} {\bibinfo  {journal} {ACS {A}pplied {M}aterials \& {I}nterfaces}\ }\textbf {\bibinfo {volume} {16}},\ \bibinfo {pages} {36315} (\bibinfo {year} {2024})}\BibitemShut {NoStop}%
\bibitem [{\citenamefont {Sun}\ \emph {et~al.}(2021)\citenamefont {Sun}, \citenamefont {Terrones},\ and\ \citenamefont {Schaak}}]{Sun2021ColloidalStructure}%
  \BibitemOpen
  \bibfield  {author} {\bibinfo {author} {\bibfnamefont {Y.}~\bibnamefont {Sun}}, \bibinfo {author} {\bibfnamefont {M.}~\bibnamefont {Terrones}},\ and\ \bibinfo {author} {\bibfnamefont {R.~E.}\ \bibnamefont {Schaak}},\ }\href {https://doi.org/10.1021/acs.accounts.1c0000} {\bibfield  {journal} {\bibinfo  {journal} {Accounts of Chemical Research}\ }\textbf {\bibinfo {volume} {54}},\ \bibinfo {pages} {1517} (\bibinfo {year} {2021})}\BibitemShut {NoStop}%
\bibitem [{\citenamefont {Bisen}\ \emph {et~al.}(2022)\citenamefont {Bisen}, \citenamefont {Atif}, \citenamefont {Mallya},\ and\ \citenamefont {Nanda}}]{Bisen2022SelfAssemblyTMDs}%
  \BibitemOpen
  \bibfield  {author} {\bibinfo {author} {\bibfnamefont {O.~Y.}\ \bibnamefont {Bisen}}, \bibinfo {author} {\bibfnamefont {S.}~\bibnamefont {Atif}}, \bibinfo {author} {\bibfnamefont {A.}~\bibnamefont {Mallya}},\ and\ \bibinfo {author} {\bibfnamefont {K.~K.}\ \bibnamefont {Nanda}},\ }\href {https://doi.org/10.1021/acsami.1c11300} {\bibfield  {journal} {\bibinfo  {journal} {ACS {A}pplied {M}aterials \& {I}nterfaces}\ }\textbf {\bibinfo {volume} {14}},\ \bibinfo {pages} {5134} (\bibinfo {year} {2022})}\BibitemShut {NoStop}%
\bibitem [{\citenamefont {Tselikov}\ \emph {et~al.}(2022)\citenamefont {Tselikov}, \citenamefont {Ermolaev}, \citenamefont {Popov}, \citenamefont {Tikhonowski}, \citenamefont {Panova}, \citenamefont {Taradin}, \citenamefont {Vyshnevyy}, \citenamefont {Syuy}, \citenamefont {Klimentov}, \citenamefont {Novikov}, \citenamefont {Evlyukhin}, \citenamefont {Kabashin}, \citenamefont {Arsenin}, \citenamefont {Novoselov},\ and\ \citenamefont {Volkov}}]{Gleb2022TMDNanoSpheres}%
  \BibitemOpen
  \bibfield  {author} {\bibinfo {author} {\bibfnamefont {G.~I.}\ \bibnamefont {Tselikov}}, \bibinfo {author} {\bibfnamefont {G.~A.}\ \bibnamefont {Ermolaev}}, \bibinfo {author} {\bibfnamefont {A.~A.}\ \bibnamefont {Popov}}, \bibinfo {author} {\bibfnamefont {G.~V.}\ \bibnamefont {Tikhonowski}}, \bibinfo {author} {\bibfnamefont {D.~A.}\ \bibnamefont {Panova}}, \bibinfo {author} {\bibfnamefont {A.~S.}\ \bibnamefont {Taradin}}, \bibinfo {author} {\bibfnamefont {A.~A.}\ \bibnamefont {Vyshnevyy}}, \bibinfo {author} {\bibfnamefont {A.~V.}\ \bibnamefont {Syuy}}, \bibinfo {author} {\bibfnamefont {S.~M.}\ \bibnamefont {Klimentov}}, \bibinfo {author} {\bibfnamefont {S.~M.}\ \bibnamefont {Novikov}}, \bibinfo {author} {\bibfnamefont {A.~B.}\ \bibnamefont {Evlyukhin}}, \bibinfo {author} {\bibfnamefont {A.~V.}\ \bibnamefont {Kabashin}}, \bibinfo {author} {\bibfnamefont {A.~V.}\ \bibnamefont {Arsenin}}, \bibinfo {author} {\bibfnamefont {K.~S.}\ \bibnamefont {Novoselov}},\ and\ \bibinfo {author} {\bibfnamefont {V.~S.}\
  \bibnamefont {Volkov}},\ }\href {https://doi.org/10.1073/pnas.2208830119} {\bibfield  {journal} {\bibinfo  {journal} {Proceedings of the National Academy of Sciences}\ }\textbf {\bibinfo {volume} {119}},\ \bibinfo {pages} {e2208830119} (\bibinfo {year} {2022})}\BibitemShut {NoStop}%
\bibitem [{\citenamefont {Nedev}\ \emph {et~al.}(2011)\citenamefont {Nedev}, \citenamefont {Urban}, \citenamefont {Lutich},\ and\ \citenamefont {Feldmann}}]{Nedev2011OFSL}%
  \BibitemOpen
  \bibfield  {author} {\bibinfo {author} {\bibfnamefont {S.}~\bibnamefont {Nedev}}, \bibinfo {author} {\bibfnamefont {A.~S.}\ \bibnamefont {Urban}}, \bibinfo {author} {\bibfnamefont {A.~A.}\ \bibnamefont {Lutich}},\ and\ \bibinfo {author} {\bibfnamefont {J.}~\bibnamefont {Feldmann}},\ }\href {https://doi.org/10.1021/nl203214n} {\bibfield  {journal} {\bibinfo  {journal} {Nano Letters}\ }\textbf {\bibinfo {volume} {11}},\ \bibinfo {pages} {5066} (\bibinfo {year} {2011})}\BibitemShut {NoStop}%
\bibitem [{\citenamefont {Staude}\ \emph {et~al.}(2013)\citenamefont {Staude}, \citenamefont {Miroshnichenko}, \citenamefont {Decker}, \citenamefont {Fofang}, \citenamefont {Liu}, \citenamefont {Gonzales}, \citenamefont {Dominguez}, \citenamefont {Luk}, \citenamefont {Neshev}, \citenamefont {Brener},\ and\ \citenamefont {Kivshar}}]{Staude2013TailoringScattering}%
  \BibitemOpen
  \bibfield  {author} {\bibinfo {author} {\bibfnamefont {I.}~\bibnamefont {Staude}}, \bibinfo {author} {\bibfnamefont {A.~E.}\ \bibnamefont {Miroshnichenko}}, \bibinfo {author} {\bibfnamefont {M.}~\bibnamefont {Decker}}, \bibinfo {author} {\bibfnamefont {N.~T.}\ \bibnamefont {Fofang}}, \bibinfo {author} {\bibfnamefont {S.}~\bibnamefont {Liu}}, \bibinfo {author} {\bibfnamefont {E.}~\bibnamefont {Gonzales}}, \bibinfo {author} {\bibfnamefont {J.}~\bibnamefont {Dominguez}}, \bibinfo {author} {\bibfnamefont {T.~S.}\ \bibnamefont {Luk}}, \bibinfo {author} {\bibfnamefont {D.~N.}\ \bibnamefont {Neshev}}, \bibinfo {author} {\bibfnamefont {I.}~\bibnamefont {Brener}},\ and\ \bibinfo {author} {\bibfnamefont {Y.}~\bibnamefont {Kivshar}},\ }\href {https://doi.org/10.1021/nn402736f} {\bibfield  {journal} {\bibinfo  {journal} {ACS Nano}\ }\textbf {\bibinfo {volume} {7}},\ \bibinfo {pages} {7824} (\bibinfo {year} {2013})}\BibitemShut {NoStop}%
\end{thebibliography}%

\end{document}